\documentclass[12pt]{elsarticle}

\usepackage{amssymb}

\usepackage{graphicx}
\usepackage{xcolor}
\usepackage{makecell}
\usepackage{array}
\usepackage{ragged2e}
\newcolumntype{P}[1]{>{\RaggedRight\hspace{0pt}}p{#1}}
\usepackage{float}
\usepackage{pifont}
\usepackage{fontawesome}
\usepackage[flushleft]{threeparttable}
\usepackage{subcaption}
\usepackage{multicol}
\usepackage{multirow}
\usepackage{tablefootnote}
\usepackage{tikz}
\usepackage{url} 
\usepackage{hyperref}
\usepackage{booktabs}
\usepackage{longtable}
\usepackage{threeparttablex}
\usepackage{afterpage}
\newcommand{\faIcon}[1]{\faicon{#1}}

\newcommand{\todo}[1]{}
\usepackage[many]{tcolorbox}    	
\usepackage{setspace}               
\definecolor{tableGray}{RGB}{243, 244, 245}
\definecolor{borderGray}{RGB}{229, 230, 233}

\newtcolorbox{boxA}{
    colback = tableGray, 
    boxrule = 0pt  
}

\definecolor{main}{HTML}{2C3E50}    
\definecolor{sub}{HTML}{F2F3F4}     

\newtcolorbox{boxB}{
    sharpish corners, 
    colback = sub, 
    boxrule = 0pt, 
    leftrule = 6pt, 
    enhanced,
    fuzzy shadow = {0pt}{-2pt}{-0.5pt}{0.5pt}{black!35}, 
    before skip = 0.2 cm, 
    after skip  = 0.2 cm  
} 

\usetikzlibrary{
    arrows.meta, 
    positioning  
}

\journal{Information and Software Technology}

\begin{document}

\begin{frontmatter}





\title {The Impact of Human Aspects on the Interactions Between Software Developers and End-Users in Software Engineering: A Systematic Literature Review}

\author[1]{Hashini Gunatilake}
\ead{hashini.gunatilake@monash.edu}

\affiliation[1]{
    organisation={Department of Software Systems and Cybersecurity, Faculty of Information Technology, Monash University},
    city={Melbourne},
    country={Australia}
}

\author[1]{John Grundy}
\ead{john.grundy@monash.edu}

\author[1]{Rashina Hoda}
\ead{rashina.hoda@monash.edu}

\author[1]{Ingo Mueller}
\ead{ingo.mueller@monash.edu}

\begin{abstract}
    \textit{Context:} Research on human aspects within the field of software engineering (SE) has been steadily gaining prominence in recent years. These human aspects have a significant impact on SE due to the inherently interactive and collaborative nature of the discipline.
    
    \textit{Objective:} In this paper, we present a systematic literature review (SLR) on human aspects affecting developer-user interactions. The objective of this SLR is to plot the current landscape of primary studies by examining the human aspects that influence developer-user interactions, their implications, interrelationships, and how existing studies address these implications.
    
    \textit{Method:} We conducted this SLR following the guidelines proposed by Kitchenham et al. We performed a comprehensive search in six digital databases, and an exhaustive backward and forward snowballing process. We selected 46 primary studies for data extraction.
    
    \textit{Results:} We identified various human aspects affecting developer-user interactions in SE, assessed their interrelationships, identified their positive impacts and mitigation strategies for negative effects. We present specific recommendations derived from the identified research gaps.
    
    \textit{Conclusion:} Our findings suggest the importance of leveraging positive effects and addressing negative effects in developer-user interactions through the implementation of effective mitigation strategies. These insights may benefit software practitioners for effective user interactions, and the recommendations proposed by this SLR may aid the research community in further human aspects related studies.
\end{abstract}


\begin{highlights}
    \item This SLR analyses 46 studies consolidating a range of human aspects that significantly influence developer-user interactions, encompassing both positive impacts and strategies for mitigating negative effects.
    
    \item The SLR findings both clarify existing knowledge and highlight research gaps, providing a set of recommendations that will influence future research in human aspects in software engineering.
\end{highlights}


\end{frontmatter}



\section{Introduction} \label{SEC:Introduction}
The cooperative and human aspects of software engineering (SE) have been studied for several decades. More recently, research on human aspects within SE such as emotions, gender, and personality, has received increasing attention \cite{pirzadeh2010humanfactors, grundy2021impact, murphy2010human}. The term ``human aspects" in SE has been broadly defined as different aspects of human involvement and influence in the SE process, spanning individual to organisational, tactical to psychological, and involving both customers and developers \cite{pirzadeh2010humanfactors}. Human aspects in software development is seen as the ingredient that ``ultimately gives a project team its soul'' \cite{brereton2007lessons}. There are various human aspects that impact SE \cite{pirzadeh2010humanfactors, hazzan2004humanaspects}. Many studies have been conducted to identify the impact of human aspects such as culture \cite{shah2012culture}, gender \cite{ribaupierre2018gender}, emotions \cite{madampe2022emotions}, human values \cite{perera2021humanvalues}, personality \cite{barroso2017personality, hidellaarachchi2022personality}, motivation \cite{unkelos2018motivation}, communication \cite{davis2006communication, anwar2011communication}, empathy \cite{levy2018empathy} and others in diverse SE contexts \cite{grundy2021addressing, capretz2014humanfactors}. These human aspects have a great influence on SE due to the inherently interactive and collaborative nature of the SE discipline \cite{pirzadeh2010humanfactors, grundy2021impact, murphy2010human}. 

Involving users in the software development process is considered useful due to their ability to influence system success \cite{abelein2013systemsuccess, abelein2015understanding}. Engaging with users in software development can enhance both the system quality and its value to users. However, some studies have highlighted drawbacks of involving users in the software development \cite{cavaye1995user, barki1994user, zeffane1998does, amoako1993user, amoako2007perceived}. These drawbacks include an increased likelihood of conflicts, negative impact on data quality, unfavourable user attitude towards system, and negative developer attitude towards users.



Motivated by the above contradictory findings and the absence of systematically accumulated knowledge in the area of the impact of developer-user interactions, we compiled primary studies that explored the influence of human aspects on these interactions within SE. In this paper, we performed a comprehensive analysis of existing studies that examined the effects of human aspects on interactions between developers and users. Through this systematic literature review (SLR), we intended to identify the different human aspects influencing these interactions, understand the implications of these human aspects, and report strategies for maximising positive effects while mitigating negative impacts. 
While our research focuses on understanding the impact of human aspects on developer-user interactions, we acknowledge the prevalence of such interactions in agile methodologies. Our study does not specifically target a single software development methodology. However, we recognise the significance of agile practices in facilitating developer-user interactions due to their iterative and customer-centric nature. Although agile practices hold significance in facilitating developer-user interactions, our study adopts a holistic approach by examining these interactions within a broader context that encompasses various software development methodologies including agile, waterfall, and other iterative approaches. By exploring interactions across different methodologies, we aim to provide insights that are applicable to a wide range of software development contexts.

We developed an SLR protocol following the guidelines provided by Kitchenham et al. \cite{Kitchenham2007Guidelines, Kitchenham2004Procedures}. After searching and filtering, we found 46 primary studies and extracted data from them. Our analysis covered various factors, including study goals, methodologies, participants, and outcomes. Through this analysis, we identified a spectrum of human aspects that influence developer-user interactions and the field of SE. Our investigation covered positive and negative effects of these human aspects, and strategies for mitigating the negative effects alongside exploring their interrelationships. Additionally, our analysis of the limitations and future work in these studies enabled us to identify key research gaps, which led to the formulation of specific recommendations.
The key contributions of this work are:
\begin{itemize}
    \item Identification of relevant human aspects and the impact of these human aspects on the interactions between software developers and users in SE context.
    \item Collation of mitigation strategies to overcome the negative effects of these human aspects. 
    \item Identification of limitations and future work of studies to inform research gaps in the area of human aspects' impact on developer-user interactions.
    \item A set of recommendations from the identified research gaps to direct future research and implications for practice.
\end{itemize}

We recognise software development as a collaborative effort that involves various roles and responsibilities, each contributing to the creation and delivery of software products. We mainly used the term `developers' to refer to programmers who are the individuals directly involved in coding activities. However, upon further examination of primary studies, we recognised that the term encompasses a broader spectrum of roles within the software development process. This includes individuals such as technical leads, head of IT department, customer support personnel, and SE students. These roles were explicitly defined as types of developers in the analysed primary studies. Additionally, we encountered other roles classified as developers, such as vendor chief architects and vendor project managers. Despite not being directly involved in coding activities, these roles were identified as developers within the primary studies. To address this complexity, we categorised these additional roles as `developer representatives'. Therefore, when we use the term `developers' in this paper, it may encompass the diverse range of professionals involved in the development life cycle.

The rest of this paper is organised as follows. Section \ref{SEC:Related Work} presents an overview of key related studies. Section \ref{SEC:Research Methodology} describes our research methodology. Section \ref{SEC:Data Analysis and Findings} describes the data synthesis and key findings of this work, and Section \ref{SEC:Discussion} discusses the identified research gaps and recommendations for future work. In Section \ref{SEC:Threats to Validity}, we present threats to validity of this SLR, followed by our conclusions in Section \ref{SEC:Conclusion}.

\section{Related Work} \label{SEC:Related Work}
Many empirical studies and reviews have been conducted on a variety of human aspects in SE. For example, the SLR conducted by Pirzadeh points out the need for conducting more research related to human aspects in the SE context \cite{pirzadeh2010humanfactors}. Several SLRs have been conducted to explore a wide range of key human aspects, such as culture \cite{fazli2017cultural, alsanoosy2020cultural}, personality \cite{barroso2017personality, cruz2011personality}, motivation \cite{beecham2008motivation, hall2009motivation}, and cognitive biases \cite{mohanani2020coginitvebias}. Additionally, SLRs have investigated the impact of various human factors on different aspects of SE, such as software quality \cite{guveyi2020softwarequality}, software development \cite{wagner2018productivity}, and system success \cite{abelein2015upi, bano2013upi}. In the next section, we outline some of these reviews. 

\textit{Culture} influences various aspects of project management, team collaboration, and the design of software products, posing challenges in cross-cultural interactions that require formal and informal mentoring, as emphasised in \cite{fazli2017cultural, alsanoosy2020cultural, marcus2000crosscurrents, casado2011culture}.
\textit{Personality} traits are extensively studied in SE, with research areas covering pair programming, education, software engineer characteristics, and team effectiveness, often using models such as Myers-Briggs type indicator and five factor model \cite{barroso2017personality, cruz2011personality}.
\textit{Motivation} studies in SE have explored factors influencing developers' motivation, demotivation, and their alignment with existing models, and have analysed motivational factors in work practices \cite{beecham2008motivation, hall2009motivation, sach2010motivation, deak2016motivation}.
\textit{Cognitive biases} in SE have been examined in the systematic mapping study conducted by Mohanani et al. The study identified cognitive biases that can influence decision-making and problem-solving in software development. The findings revealed a wide range of cognitive biases prevalent in SE, and helped to identify techniques to mitigate cognitive biases in SE activities \cite{mohanani2020coginitvebias}.

In a SLR on the \textit{impact of human factors on software quality}, personal factors were identified as more important than interpersonal or organisational factors, emphasising the significance of addressing human aspects to enhance software quality and project success \cite{guveyi2020softwarequality}. The SLR by Wagner and Ruhe on \textit{productivity factors in software development} identified a range of factors, categorising them into technical and soft aspects, with soft factors including developer skills, team size, team cohesion, experience, and respect \cite{wagner2018productivity}.

Additionally, a SLR by Abelein et al. focused on \textit{user participation and involvement (UPI)} in system success. This study found that human aspects positively affect system success, and UPI has a positive impact on user satisfaction and system use. The researchers recommended further empirical research to explore UPI in specific contexts and its relationship with various contextual factors \cite{abelein2015upi}. Another SLR examined the \textit{impact of user involvement} on system success, reporting an overall positive effect of user involvement on system success. This study emphasised that the relationship between user involvement and system success is complex, influenced by various factors and conditions in the development process \cite{bano2013upi}.

While there have been several SLRs conducted in the field of SE, most of them have primarily focused on individual human aspects such as personality, motivation, cognitive bias, and their influence within specific SE-related processes, including but not limited to cultural influences on collaborative work in SE teams, and cultural effects on requirements engineering (RE) activities. Additionally, there are SLRs that explored the effects of various human factors on different aspects of SE, such as software quality, software development, system success, and RE. Several studies report investigations and surveys of stakeholder/customer interactions and suggest gaps in research around stakeholder/customer interactions in SE \cite{saiedian2000requirements,whitehead2007collaboration,sauvola2015towards,fabijan2015customer}. Critically, there appears to be no SLR that specifically investigates the impact of human aspects on the interactions between software developers and users. Given the vital importance of this relationship in SE, it makes sense to accumulate the collective knowledge on this topic through a dedicated SLR, so SE researchers and practitioners can benefit from learning about the state of the art.



\section{Research Methodology} \label{SEC:Research Methodology}
\begin{figure} [t]
    \centering
    \includegraphics[scale = 0.37]{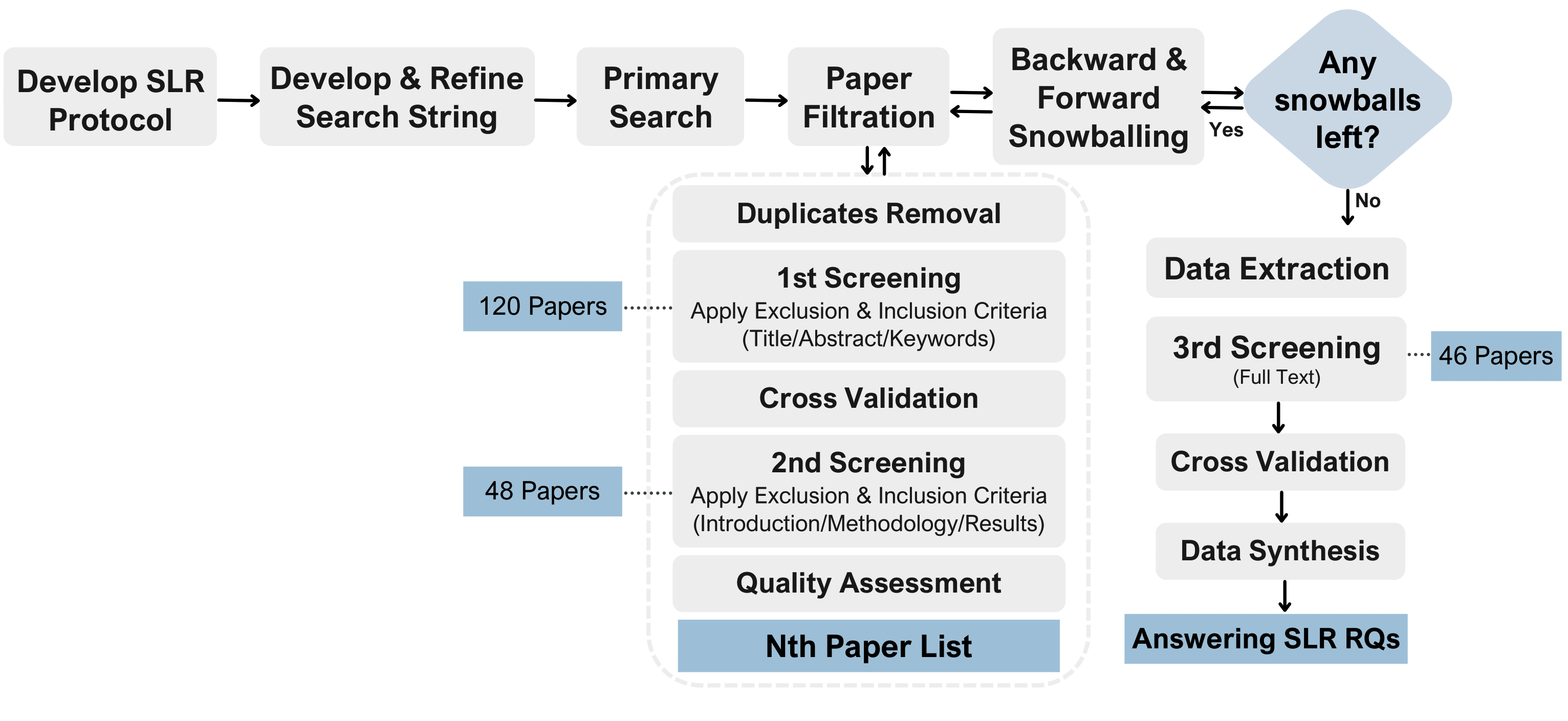}
    \caption{Overview of SLR Methodology}
    \label{FIG:Overview of SLR Methodology}
\end{figure}

We conducted a SLR to systematically analyse the existing primary studies pertaining to the influence of human aspects on the interactions between software developers and end-users during SE activities. 
This SLR adheres to the well-defined SLR methodology introduced by Kitchenham and Charters' guideline \cite{Kitchenham2007Guidelines} and Kitchenhams' procedures \cite{Kitchenham2004Procedures}. 
The first author developed a review protocol, outlining each step of the SLR process. 
The next steps of searching primary studies, study filtration, quality assessment, and data extraction was carried out with consultation of the other authors experienced in SLRs and tertiary studies in the SE context. After quality assessment, 46 highly relevant primary studies were identified and information was extracted. These 46 studies were analysed to explore diverse human aspects affecting developer-user interactions, by evaluating the study methodologies, proposed solutions and the effects of relevant human aspects. For data synthesis, we employed a widely recognised meta-analysis technique as outlined in \cite{pigott2020metaanalysis}.
An overview of SLR methodology is illustrated in Figure \ref{FIG:Overview of SLR Methodology}.

In conducting our SLR, we made deliberate methodological decisions to ensure the rigour and comprehensiveness of our study. One aspect we carefully considered was the initiation of our study based on previous SLRs on human aspects of SE (see Section \ref{SEC:Related Work}). While this approach could have potentially leveraged existing literature and saved time and resources, we opted for a different approach for several reasons.
Firstly, our objective was to investigate the impact of human aspects on developer-user interactions in SE. 
\textcolor{black}{Prior to initiating our SLR, we conducted an extensive search across multiple digital databases (e.g., IEEEXplore, ACM DL, Wiley, TFO, SpringerLink, ScienceDirect) and other sources (e.g., Google Scholar) to identify any secondary or tertiary studies investigating the impact of human aspects on developer-user interactions. Through this search process, we thoroughly reviewed the available literature and found no existing secondary or tertiary studies addressing this specific topic.} Despite the existence of secondary and tertiary studies focusing on various aspects of SE, including individual human aspects and their impact on different facets of SE, there were no comprehensive studies specifically examining the impact of human aspects on developer-user interactions. Therefore, we chose to conduct our own primary search to provide a comprehensive overview of the current state of research in this specific area. \textcolor{black}{This decision was based on the absence of relevant literature in the secondary and tertiary sources, as well as the comprehensive examination of primary studies.}
Secondly, while tertiary study approaches can be valuable in synthesising existing literature, they are inherently limited by the scope and coverage of the primary and secondary studies included in the analysis. By conducting our own primary search, we were able to directly access and evaluate a diverse range of studies, thereby reducing the risk of overlooking relevant research or biases inherent in previous reviews. Additionally, our decision aligns with the iterative and rigorous nature of SLRs, where each step, from search strategy development to study selection and synthesis, is carefully conducted to ensure methodological rigour and transparency.

\subsection{Research Questions}
We initiated our study by formulating a set of research questions (RQs) based on the framework proposed by Petticrew and Roberts \cite{Petticrew2008Systematic}. This framework, known as PICOC (population, intervention, comparison, outcomes, and context), was also outlined in Kitchenham and Charters' guideline \cite{Kitchenham2007Guidelines}. We identified PICOC concepts related to this SLR as specified in Table \ref{TAB:Table PICOC}.  

\begin{table}[htbp]
    \caption{PICOC for Research Questions}
    \label{TAB:Table PICOC}
    \scriptsize
    \begin{tabular}{ p{0.2\linewidth}  p{0.7\linewidth }}
        \toprule
        \textbf{Population} & Software developers and end-users \\
        \textbf{Intervention} & Human aspects of software developers and end-users \\
        \textbf{Comparison} & N/A \\
        \textbf{Outcomes} & Effects of human aspects on the interactions between software developers and end-users in SE context \\
        & Relationships among human aspects\\
        \textbf{Context} & Software engineering\\
        & Interactions between developers and end-users \\
        \bottomrule
    \end{tabular}
\end{table}

\textbf{RQ1. What are the objectives and methodological approaches for exploring the impact of human aspects on the interactions between software developers and end-users in SE?}

The first RQ explores the primary goals and objectives of each study, and the methodologies employed by researchers to identify the impact of human aspects on the interactions between software developers and end-users in the SE context. This RQ also examines the publication trends, roles and functions of study participants, and the nature of interactions among them.

\textbf{RQ2. What are human aspects that influence the interactions between developers and end-users?}

The second RQ investigates the human aspects of developers and end-users that have been explored to date in the SE field. It also explores the proposed solutions to address the impact of human aspects in SE, and how these solutions have been evaluated. Further this RQ analyses the positive and negative relationships among human aspects explored in the primary studies.

\textbf{RQ3. How do the identified human aspects influence the interactions between developers and end-users?}

The third RQ analyses the impact of the identified human aspects on the interactions between software developers and end-users in the SE field. This RQ aims to consolidate understanding of the benefits of promoting human aspects that have a positive impact and to explore approaches that can be employed to mitigate the negative effects of human aspects.

\textbf{RQ4. What are the identified limitations and future work?}

The fourth RQ examines the limitations and research gaps that were identified in the primary studies included in the review. This involves understanding the constraints or shortcomings of the existing research as well as the areas where further investigation is recommended by the authors of these studies.

\subsection{Search Strategy} \label{SEC:Search String Formulation and Refining}


We identified key search terms based on the PICOC concepts that were used to develop our SLR RQs (see Table \ref{TAB:Table PICOC}). Initial search query for primary search was developed using the key search terms presented in Table \ref{TAB:Table Key Search Terms}. To ensure comprehensive coverage, we defined alternative search terms by identifying the synonyms of key search terms frequently used in SE (see Table \ref{TAB:Table Alternative Search Terms}). We combined these key and alternative search terms using Boolean AND and OR operators to formulate the search query. 

\begin{table}[htbp]
    \scriptsize
    \caption{Key Search Terms}
    \label{TAB:Table Key Search Terms}
    \begin{tabular}{ p{0.17\linewidth}  p{0.73\linewidth }}
        \toprule
        \textbf{Concept} & \textbf{Key Search Term} \\
        \midrule
        Population & Software developers and end-users \\
        Intervention & Human aspects\\
        Outcomes & Effects of human aspects\\
        Context &  Interactions between developers and end-users \\
        \bottomrule
    \end{tabular}
\end{table}

\begin{table}[htbp]
    \scriptsize
    \caption{Alternative Search Terms}
    \label{TAB:Table Alternative Search Terms}
    \begin{tabular}{ p{0.25\linewidth}  p{0.7\linewidth }}
        \toprule
        \textbf{Key Search Term} & \textbf{Alternative Search Terms} \\
        \midrule
        End-Users & End-Customers/Software-Users/Users/Customers/Clients  \\
         Software Developers & Developers/Software-Engineers/Coders/Programmers\\
        Human Aspects & Human-Factors/Human-Issues/Human-Influences\\
        Interaction & Relationship/Involvement/Association/Participation\\
        \bottomrule
    \end{tabular}
\end{table}

Our search procedure was performed in two ways: database search (primary search) and snowballing (secondary search). Database search was conducted using the search engines of digital databases including IEEE Xplore, ACM Digital Library (ACM DL), Wiley Online Library (Wiley), Taylor and Francis Online (TFO), SpringerLink, and ScienceDirect. These databases were selected as they contain the most high quality, peer-reviewed papers in SE and Computer Science, and also based on the online databases subscribed by the Monash University library under the ``Computer Science'' category. Our search was not time bound as the goal was to identify all the studies which fit into the given criteria. Our primary search includes papers till April 2023. Database filters stated in Table \ref{TAB:Digital Database Filters} were used to select the most relevant primary studies. Thirteen highly relevant studies were selected from database search. The breakdown of papers in the database search are detailed in Table \ref{TAB:Primary Search Paper Breakdown}.  

\begin{table} [htbp]
    \centering
     \begin{scriptsize}   
    \caption{Digital Database (DB) Filters in Primary Search}
    \label{TAB:Digital Database Filters}
    \begin{threeparttable}
    \begin{tabular}{lllll}
        \toprule
         \textbf{Digital DB} & \textbf{Search Type} & \textbf{Search Fields} & \textbf{Article Type} & \textbf{Subjects}\\
         \midrule
         
         IEEE Xplore & Command & All metadata & Conferences, Journals, & - \\
         &  & & Early access articles & \\
          
         ACM DL & Advanced & Abstract & Research article & - \\
         Wiley & Advanced & Abstract & Journals & CompSci*\\
         TFO & Advanced & Abstract & Article & CompSci\\
         SpringerLink & Basic & Full-text* & Article, Conference paper, & - \\
         & & & Conference proceeding & \\
         ScienceDirect & Advanced &  Title, abstract or & Research articles & - \\
         & & author-specified & &\\
          & & keywords & &\\

         \bottomrule
    \end{tabular}
        \begin{tablenotes}
            \item * Search fields cannot be restricted in basic search, CompSci: Computer Science.
        \end{tablenotes}
    \end{threeparttable}
    \end{scriptsize}   
\end{table}

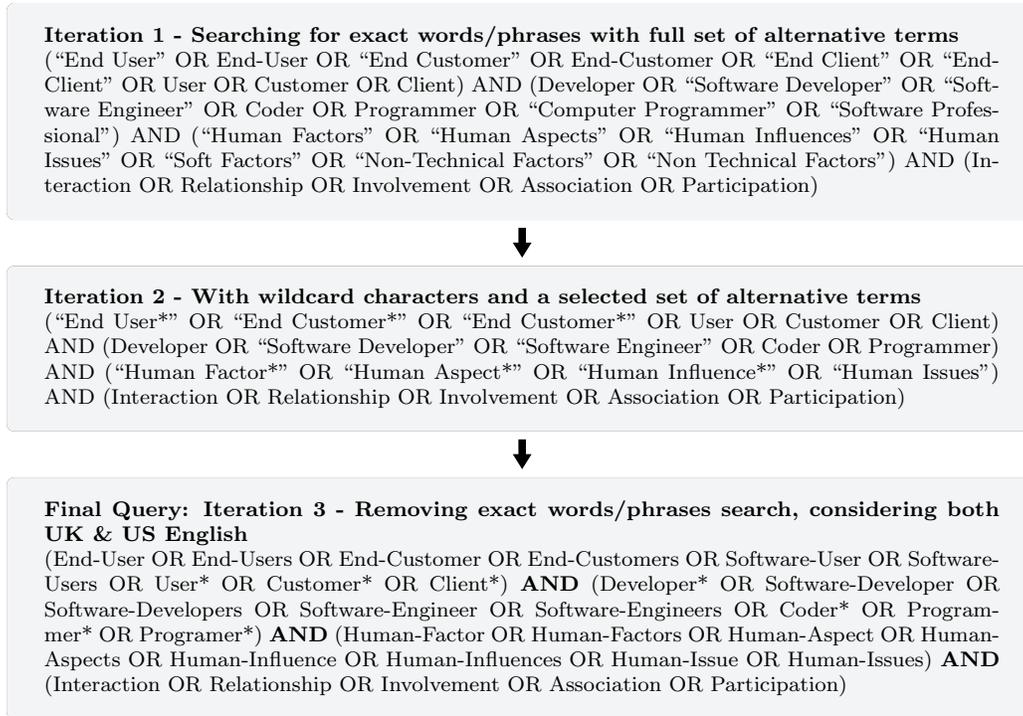
\begin{figure}[t]
    \centering

    \begin{scriptsize}   
    \begin{tikzpicture}[node distance=0.4cm]
    \node(Node1){
        \begin{boxA}
        \textbf{Iteration 1 - Searching for exact words/phrases with full set of alternative terms}\\ 
        (``End User'' OR End-User OR ``End Customer'' OR End-Customer OR ``End Client'' OR ``End-Client'' OR User OR Customer OR Client) AND (Developer OR ``Software Developer'' OR ``Software Engineer'' OR Coder OR Programmer OR ``Computer Programmer'' OR ``Software Professional'') AND (``Human Factors'' OR ``Human Aspects'' OR ``Human Influences'' OR ``Human Issues'' OR ``Soft Factors'' OR ``Non-Technical Factors'' OR ``Non Technical Factors'') AND (Interaction OR Relationship OR Involvement OR Association OR Participation)
        \end{boxA}
   };
   
   \node(Node2) [below=of Node1] {  
        \begin{boxA}
        \textbf{Iteration 2 - With wildcard characters and a selected set of alternative terms}\\
        (``End User*'' OR ``End Customer*'' OR ``End Customer*'' OR User OR Customer OR Client) AND (Developer OR ``Software Developer'' OR ``Software Engineer'' OR Coder OR Programmer) AND (``Human Factor*'' OR ``Human Aspect*'' OR ``Human Influence*'' OR ``Human Issues'') AND (Interaction OR Relationship OR Involvement OR Association OR Participation)
        \end{boxA}
    };
    
    \draw[black, line width=1mm,-{Triangle[angle=60:1pt 2]}] (Node1) -- (Node2);
    
    \node(Node3) [below=of Node2] {  
        \begin{boxA}    
        \textbf{Final Query: Iteration 3 - Removing exact words/phrases search, considering both UK \& US English}\\
        (End-User OR End-Users OR End-Customer OR End-Customers OR Software-User OR Software-Users OR User* OR Customer* OR Client*) \textbf{AND} (Developer* OR Software-Developer OR Software-Developers OR Software-Engineer OR Software-Engineers OR Coder* OR Programmer* OR Programer*) \textbf{AND} (Human-Factor OR Human-Factors OR Human-Aspect OR Human-Aspects OR Human-Influence OR Human-Influences OR Human-Issue OR Human-Issues) \textbf{AND} (Interaction OR Relationship OR Involvement OR Association OR Participation)
        \end{boxA}
    };
    
    \draw[black, line width=1mm,-{Triangle[angle=60:1pt 2]}] (Node2) -- (Node3);

    \end{tikzpicture}
    \caption{Search String Formulation for Primary Search}
    \label{FIG:Figure search String Formulation}
    \end{scriptsize}
\end{figure}

We created different versions of the search query, by modifying it using standard search techniques, and executed these different versions in each database to determine the most effective search query.
These techniques include wildcards(*), stemming, searching for exact words/phrases, and considering UK and US English variations. The most effective search query was determined by assessing the relevance of retrieved papers, with the query resulting in the highest number of relevant papers being identified as such. We tried to use the same set of search terms for all the databases but we were able to omit some terms based on the database search rules. We used the term ``human aspects" and its synonyms (see Table \ref{TAB:Table Alternative Search Terms}) in the search query without specifying individual human aspects (e.g., emotions, culture). This approach aimed to prevent bias towards any particular aspect, and it was also necessitated by database constraints that prevented employing all existing human aspects in a single search query. While this approach may have limited the number of papers retrieved during the primary search, its intention was to ensure unbiased results.
The formulation and refinement process is outlined in Figure \ref{FIG:Figure search String Formulation}. The search strings used for each database are accessible online in our supplementary information package. \footnote{https://github.com/Hashini-G/SupplementaryInfoPackage-SLR}

\begin{table} [ht]
    \centering
    \scriptsize
    \caption{Paper Breakdown of Primary Search }
    \label{TAB:Primary Search Paper Breakdown}
    \begin{threeparttable}
    \begin{tabular}{llllll}
        \toprule
         Digital DB & Initial & After dupli- & 1st & 2nd & 3rd \\
         & Count & cates rmv.  & scr. & scr. & scr. \\
         
         \midrule
         
         IEEE Xplore & 244 & 243 & 35 & 7 & 6\\
         ACM DL & 102 & 98 & 6 & 1 & 1\\
         Wiley & 14 & 14 & 5 & 1 & 1\\
         TFO & 71 & 71 & 4 & 0 & 0\\
         SpringerLink & 9238 & 8981 & 7 & 5 & 4\\
         ScienceDirect & 29 & 23 & 5 & 1 & 1\\
         \textbf{Total} & \textbf{9698} & \textbf{9430} & \textbf{62} & \textbf{15} & \textbf{13} \\
         
         \bottomrule
    \end{tabular}
    \begin{tablenotes}
        \item rmv: removal, scr: screening
    \end{tablenotes}
     \end{threeparttable}
\end{table}

The secondary search process was performed by adopting an iterative and exhaustive backward and forward snowballing approach \cite{wohlin2014snowballing} after filtering primary search results. Backward and forward snowballing was conducted until there were no snowballs left. This involved checking the references and citations of each newly included study to identify additional relevant papers. We employed this snowballing approach due to two main reasons. Firstly, individual human aspects were not included in the initial search query due to the multitude of aspects and there was no way to include all these aspects in the search query due to database limitations. Instead of being confined by database search capabilities, snowballing enabled us to cast a wider net by following citation trails across various sources, thereby capturing relevant literature. Secondly, the absence of a proper taxonomy for human aspects posed another challenge. Without a clear taxonomy, it was difficult to ensure comprehensive inclusion of all the human aspects and accurately capture the breadth of relevant literature. By adopting the snowballing approach, we were able to explore diverse human aspects beyond predefined keywords, by following leads from existing literature. This flexibility enabled us to discover relevant studies that might address aspects not covered by conventional search terms, thus enhancing the comprehensiveness of our review.
In addition, we employed another snowballing technique of searching in specific venues \cite{wohlin2014snowballing}. We specifically searched in the CHASE conference \footnote{Cooperative and Human Aspects of Software Engineering (CHASE) operated as a workshop until 2020, and since 2021, it has functioned as a working conference, co-located with the International Conference on Software Engineering (ICSE).} due to its strong alignment with our SLR focus. We considered CHASE papers from 2008 to 2023, subjecting them to the same filtration process. Overall, 33 highly relevant studies were selected from snowballing (see Table \ref{TAB:Secondary Search Paper Breakdown}) resulting in 46 total papers from both search stages. 

\begin{table}[htbp]
    \centering
    \scriptsize
    \caption{Paper Breakdown of Secondary Search }
    \label{TAB:Secondary Search Paper Breakdown}
    \begin{threeparttable}
    \begin{tabular}{lllllll}
        \toprule
         Round & Backw. & Forward & Round & 1st & 2nd & 3rd \\
         & Count & Count & Total & scr. & scr. & scr.  \\
         \midrule
         
         R1 & 454 & 175 & 629 & 21 & 14 & 14\\
         R2 & 441 & 833 & 1274 & 17 & 9 & 9\\
         R3 & 281 & 447 & 728 & 6 & 4 & 4\\
         R4 & 58 & 721 & 779 & 7 & 4 & 4\\
         R5 & 117 & 115 & 232 & 1 & 0 & 0\\
         CHASE R0* & - & - & 295 & 6 & 2 & 2\\
         CHASE R1 & 46 & 34 & 80 & 0 & 0 & 0\\
         \textbf{Total} & \textbf{1397} & \textbf{2325} & \textbf{4017} & \textbf{58} & \textbf{33} & \textbf{33} \\
         \bottomrule
    \end{tabular}
    \begin{tablenotes}
        \item rmv: removal, scr: screening
        \item *R0: Total count from CHASE 2008 to 2023
    \end{tablenotes}
     \end{threeparttable}
\end{table}

\begin{figure} [t]
    \centering
    \includegraphics[width = \textwidth]{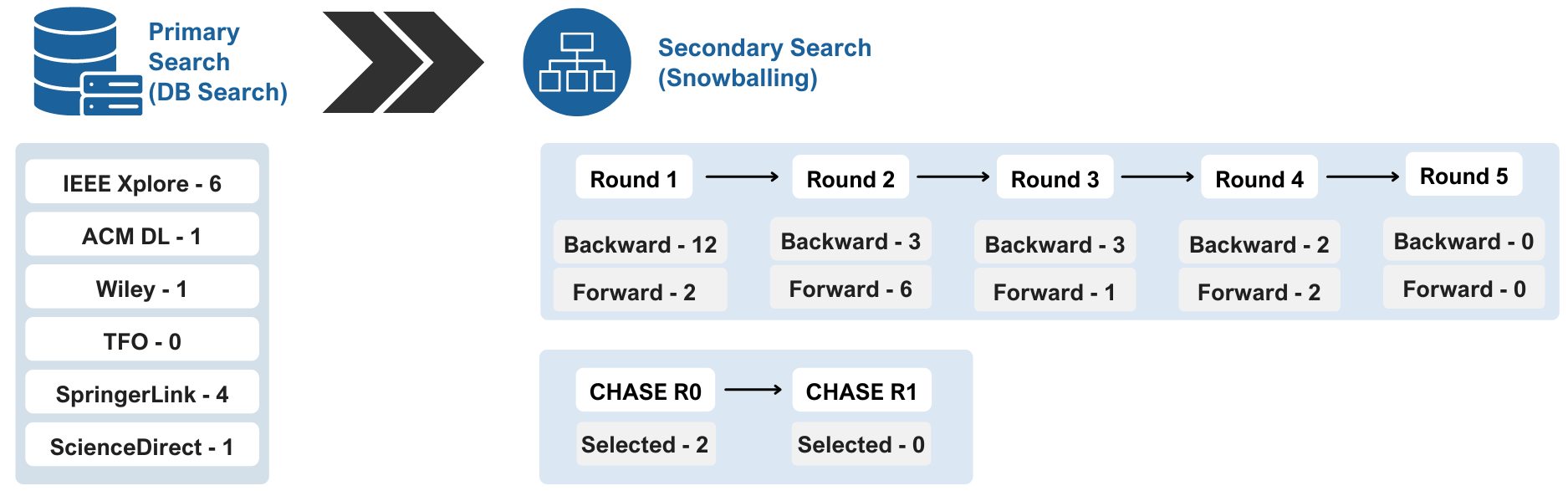}
    \caption{Overview of Paper Selection Process}
    \label{FIG:Overview of Paper Selection Process}
\end{figure}

\textcolor{black}{Figure \ref{FIG:Overview of Paper Selection Process} illustrates the changes in the selected paper sample across multiple stages until the final selection of papers is acquired.} \textcolor{black}{In Appendix \ref{Appendix:List of included papers}, we have included the papers identified through snowballing, demonstrating the breadth of human aspects covered in our study. These aspects encompass a wide range of factors such as communication, collaboration, empathy, motivation, perception, culture, emotions, challenges, performance, coordination, human values, education, personality, engagement, and cognitive style. By leveraging snowballing, we were able to uncover studies that delve into these nuanced aspects, which might have been overlooked by conventional search terms alone. This approach not only enhanced the comprehensiveness of our review but also provided valuable insights into the multifaceted nature of developer-user interactions.}

\subsection{Paper Filtration Process}

\subsubsection{Inclusion and Exclusion Criteria}
Our study employed a rigorous filtration process guided by predefined inclusion and exclusion criteria. We formulated seven inclusion criteria and six exclusion criteria (see Table \ref{TAB:Inclusion Criteria}). These criteria were established during the development of our SLR protocol and refined during the paper filtration process to uphold the integrity of our paper selection process. The inclusion and exclusion criteria were applied to all the papers to select the most relevant studies. 

\begin{table} [h!]
    \centering
    \scriptsize
    \caption{Inclusion and Exclusion Criteria}
    \label{TAB:Inclusion Criteria}
    \begin{tabular}{ p{0.05\linewidth}  p{0.9\linewidth }}
        \toprule
        \textbf{ID} & \textbf{Inclusion Criterion}\\
        \midrule
        I1 & Full text papers published as journal articles, conference papers or workshop articles that comply with search terms defined in Table \ref{TAB:Table Alternative Search Terms}.\\
        I2 & Papers used in academia (Literature references).\\
        I3 & Papers about human aspects of developers and/or end-users.\\
        I4 & Papers about challenges or gaps between developers and users.\\
        I5 & Papers about human aspects of intermediaries who facilitate developer-user interactions.\\
        I6 & Papers based on expert opinion or literature analysis that include guidelines/tools/ frameworks to improve the interactions between developers and users. \\
        I7 & The most complete version of the paper is considered in a case of duplicate articles from the same study.\\
%
        \toprule
        \textbf{ID} & \textbf{Exclusion Criterion}\\
        \midrule
        E1 & Grey literature (theses, unpublished and incomplete work), posters, books, editorial, secondary/tertiary/ review studies (SLR/SMS), discussions and keynotes.\\
        E2 & Short papers where page count is less than three pages, irrelevant and low quality papers that do not contain sufficient information to extract.\\
        E3 & Papers based only on authors' personal views without supporting data.\\
        E4 & Papers when the full version is unavailable due to university subscription restrictions.\\
        E5 & Papers about developer and end-user interactions but not focused on human or technical aspects.\\
        E6 & Papers that are not fully-written in English.\\
        \bottomrule
    \end{tabular}
\end{table}

\subsubsection{Filtering of the Papers} \label{SEC:Filtering of the Papers}

Our paper filtration process involved three screening phases. Prior to this process, we downloaded search results from databases and removed duplicates using EndNote \footnote{EndNote Reference Management Software}. We used Google Sheets to deduplicate SpringerLink results due to the lack of an EndNote-compatible export format. We first applied exclusion criteria based on the title, abstract, and keywords. Subsequently, we applied our inclusion criteria resulting in 62 papers. During this process, we categorised papers into three main groups: \textit{Relevant} for papers that directly matched our RQs and inclusion criteria, \textit{Somewhat relevant} for papers that appeared pertinent but needed closer examination, and \textit{Irrelevant} for papers that didn't fit our RQs and matched exclusion criteria. There were several reasons for excluding papers in our first screening phase. Many papers were omitted because they did not meet our inclusion criteria or met our exclusion criteria. For instance, some papers were secondary or tertiary studies, which did not meet our requirement for primary research. Additionally, we excluded papers not written in English, those with fewer than three pages, and those discussing human aspects unrelated to developers, users, or their intermediaries, adhering to our exclusion and inclusion criteria. Further, we encountered conference summaries, particularly related to CHASE, which were not suitable for our study. Additionally, several papers focused on topics such as gamification, UX, usability, ergonomics, and end-user related SE did not address the human aspects of developer-user interactions, leading to their exclusion from our final analysis.

In the second screening, we conducted a preliminary analysis of all \textit{relevant} and \textit{somewhat relevant} papers by reading title, keywords, abstract, conclusion, and skimming introduction, methodology, and results. We then applied our exclusion and inclusion criteria. During this filtration phase, we conducted a more in-depth examination of the papers. Papers were primarily omitted due to either a mismatch in research focus or a mismatch in paper type. Regarding the mismatch in research focus, we found several papers that did not address human aspects or challenges related to developer-user interactions, as well as papers discussing human aspects of other roles in SE besides developers and users. In terms of paper type, we identified several experiential papers (based on expert opinion) and papers solely providing literature reviews without offering clear outcomes to enhance developer-user interactions. All these papers were excluded adhering to the inclusion and exclusion criteria of our SLR.
After this phase, we assessed the quality of selected papers, as described in Section \ref{SEC:Quality Assessment}. To ensure a rigorous and unbiased filtration process, a total of 12 papers were distributed among other co-authors for independent evaluation. During this cross-validation, a high degree of similarity was observed. All discrepancies were thoroughly discussed and resolved during our regular fortnightly meetings. We identified 15 highly relevant papers in this phase.

The third screening was performed during the data extraction. We read full papers and found a few papers that did not adequately address the RQ-based fields in our data extraction form, despite their initial classification as relevant. Our final paper count after this phase was 13 papers (see Table \ref{TAB:Primary Search Paper Breakdown}). After filtering the papers from primary search, we performed snowballing as described in Section \ref{SEC:Search String Formulation and Refining} to identify additional relevant papers. We applied the same screening phases to these papers and found 33 relevant papers (see Table \ref{TAB:Secondary Search Paper Breakdown}).

\subsection{Quality Assessment} \label{SEC:Quality Assessment}
All selected papers were assessed for quality at the end of the second screening phase of the paper filtration process. We devised a scoring mechanism that ranged from one to five, encompassing categories such as \textit{very poor (1)}, \textit{inadequate (2)}, \textit{moderate (3)}, \textit{good (4)}, and \textit{excellent (5)}. Quality evaluation was systematically applied to all the selected papers using a set of eight distinct criteria  detailed in Table \ref{TAB:Quality Assessment Criteria}. Each criterion was assigned a weight and consequently each paper received a rating between one and five, indicative of its quality ranging from ``very poor" to ``excellent". Based on this scoring mechanism, the highest attainable quality score for a primary study was set at five. Papers were classified as ``low quality" if their average quality score was below two. During the second screening phase, we identified three ``low quality" papers and excluded them from the final paper set.


\begin{table}[htbp]
    \centering
    \scriptsize
    \caption{Quality Assessment Criteria}
    \label{TAB:Quality Assessment Criteria}
    \begin{tabular}{ p{0.05\linewidth}  p{0.9\linewidth}}
        \toprule
        \textbf{ID} & \textbf{Quality Criterion}\\
        \midrule
         QC1 & Is the paper highly applicable to the proposed SLR?\\
         QC2 & Is there a clear statement of the aim of the research?\\
         QC3 & Is there a review of key past work?\\
         QC4 & Is there a clear research methodology which aligns with key research questions of the study?\\
         QC5 & Does paper provide sufficient information on data collection and data analysis of the research?\\
         QC6 & Are the findings of the research clearly stated and supported by the research questions?\\
         QC7 & Does the paper provide limitations, summary and future work of the research?\\
         QC8 & Is the paper published in a reputable venue?\\
         \bottomrule
    \end{tabular}
\end{table}

\subsection{Data Extraction Strategy} \label{SEC:Data Extraction Strategy}
We employed a Google Form for data extraction to ensure consistency in the extracted data from each research paper. 
The data extraction form comprised 47 questions, divided into distinct sections such as publication details, key areas of the study, research methodology, research gaps, limitations \& future work, and research findings. Detailed data extraction criteria are available in our online supplementary information package. \footnote{https://github.com/Hashini-G/SupplementaryInfoPackage-SLR} 
Prior to the actual data extraction, we refined the question flow and form structure by conducting four pilot data extractions using papers from each database. The second author reviewed the data extraction form and pilot tests. Upon discussing and finalising the data extraction form, the first author extracted data from four papers, each selected from different databases. The same set of papers was independently extracted by the second author for cross-validation purposes. A very high degree of similarity was observed in the extracted data during this cross-validation phase. Any discrepancies were thoroughly discussed and resolved ensuring an unbiased data extraction process. The first author undertook the data extraction process for the remaining 42 papers, under the close supervision and guidance of the other authors. In addition, regular fortnightly meetings were held, during which the first author discussed the data extraction process with other authors, allowing for peer review and feedback.



\section{Data Analysis and Findings} \label{SEC:Data Analysis and Findings}
Data was extracted from all 46 primary studies, encompassing a diverse range of data types, including qualitative e.g. details of the study design, study findings, limitations; quantitative e.g. number of studies with evaluation details, number of studies that adopted qualitative data analysis techniques, number of studies conducted in industry setting; and mixed e.g. summarising statistical analyses and qualitative findings supporting specific conclusions from the primary studies. Our analysis involved the utilisation meta-analysis technique \cite{pigott2020metaanalysis, shelby2008understanding, ahn2018introduction} and visualisation tools. Meta-analysis, which involves statistical techniques and thematic analysis for synthesising findings from multiple studies, guided our data analysis process. Qualitative data analysis commenced by identifying patterns or themes within the data and interpreting their meanings. Subsequently, we systematically categorised and organised extracted data pertaining to related research questions based on their content or meaning. This approach allowed answering research questions by identifying common themes or ideas across the primary studies. Quantitative analysis involved summarising and describing the data using measures such as frequencies or percentages. It enabled the aggregation of results from multiple studies to derive overall estimates or effects sizes. In cases involving mixed data types, we integrated both quantitative and qualitative analysis to draw comprehensive conclusions. Visualisations, such as graphs, charts and tables, were employed to present the integrated findings in a clear and concise manner, allowing for comparisons and interpretations across various data types. The first author synthesised the data under the guidance of other authors, and visualised the findings employing various graphs, figures, and tables. The selected primary studies for this SLR are listed in \ref{Appendix:List of included papers} and detailed data synthesis spreadsheet is available in our online supplementary information package. \footnote{https://github.com/Hashini-G/SupplementaryInfoPackage-SLR}

We identified a total of 22 journal papers and 24 conference/workshop papers within the pool of 46 selected studies. We identified primary studies that spanned from 1986 to 2022, as we did not impose a specific time constraints. Since the paper list was gathered till April 2023, there may still be a few papers published after our search.


In our analysis, we observed that the subject of the majority of primary studies was related to the software industry, accounting for 80.4\% of the total. Additionally, 8.7\% of the studies were primarily focused on academia, while 10.9\% of the studies examined both industry and academia. 
We identified ten studies that conducted region-specific experiments, such as those exclusively involving Swedish software engineers, Turkish defence projects, Malaysian small and medium enterprises (SMEs), and experiments conducted within an Australian context (SBB12, SBF11, WILEY01, IEEE05). 
We found a diverse set of application domains within the studies, including healthcare, sales and manufacturing, telecommunications, education, technology, human resource management, retail sector, and more (see Table \ref{TAB:Categorisation by Application Domain}). The domains with the highest number of studies were education, healthcare, and sales \& manufacturing.

Some papers provided details about the SE phases under consideration (see Table \ref{TAB: Categorisation of Studies by SE Phases}). The requirement elicitation phase was the most frequently studied (13 papers). Design, implementation, and maintenance phases were explored by several studies (6, 6, and 7 papers, respectively), while testing was less common (4 papers). Several studies identified the SE phases most impacted by human aspects. Four papers highlighted requirement elicitation (SPR01, SBB04, SBB15, SBF02), two pointed to design and implementation (IEEE01, SBF01), and one focused on maintenance (WILEY01).

\begin{figure} [t]
    \centering
    \includegraphics[width=\textwidth]{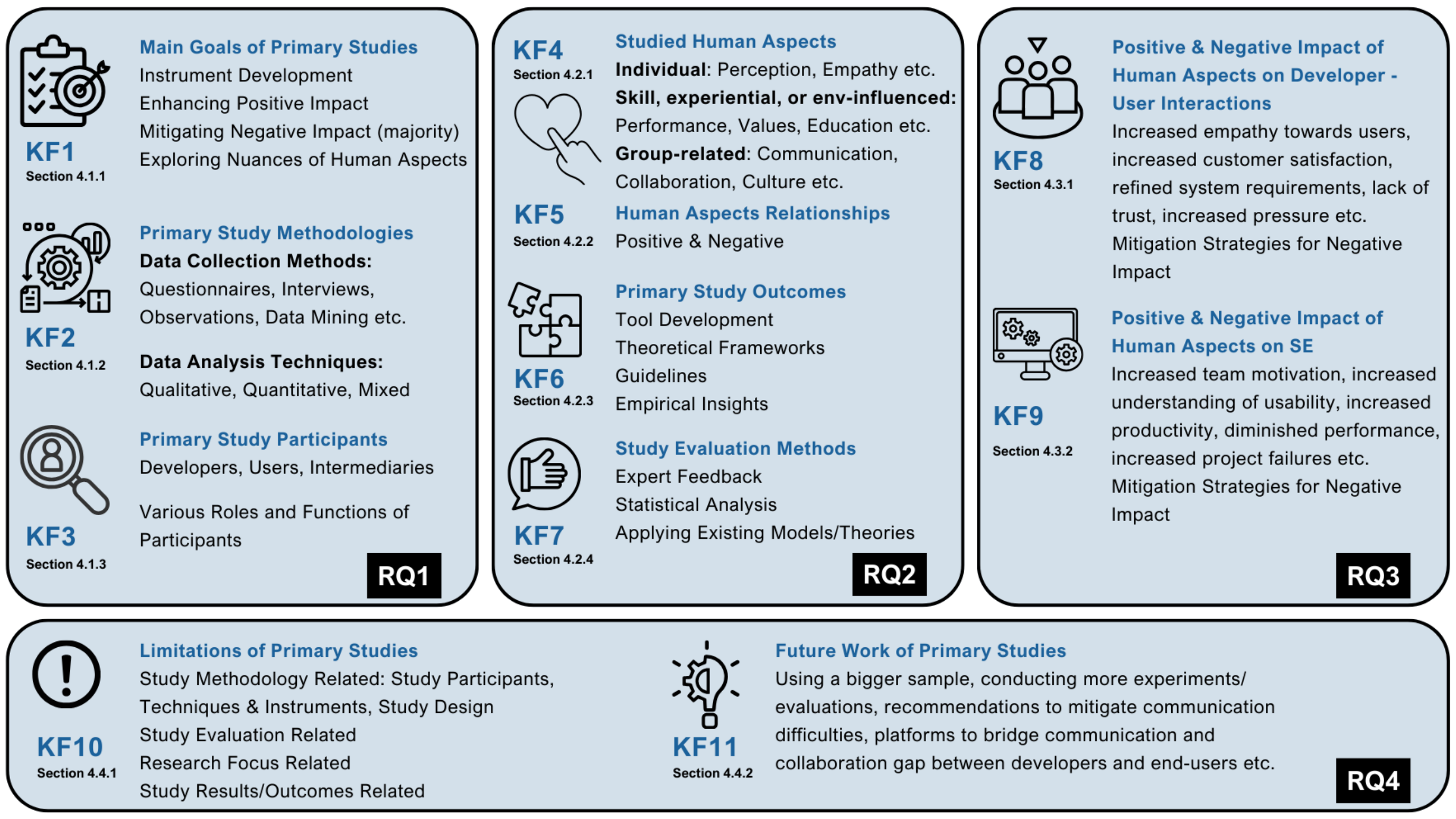}
    \caption{Summary of the Key Findings of this SLR}
    \label{FIG:Key Findings}
\end{figure}

In the following subsections, we answer each of our research questions. Figure \ref{FIG:Key Findings} provides an overview of our key findings.

\begin{table} [t]
    \centering
    \begin{threeparttable}
    \scriptsize
    \caption{Primary Study Categorisation by Application Domain}
    \label{TAB:Categorisation by Application Domain}
   
    \begin{tabular}{p{0.15\linewidth}  p{0.35\linewidth} p{0.15\linewidth}  p{0.2\linewidth}}
        \toprule
         \textbf{Application Domain} & \textbf{Studies} & \textbf{Application Domain} & \textbf{Studies} \\
         \midrule
         
         Education & SBB02, SBB05, SBB16, SBF09 & HRM* & SBB13, SBF01\\
         Healthcare &  SD01, IEEE06, SBB07, SBF03 & Insurance & SPR02\\
         SM* & SBB02, SBB06, SBB05, SBB11 & Financial & IEEE06\\
         Tech* & SPR02, SBF03 & Energy & SBF03\\
         BMC* & SPR02, IEEE06 & Defense & SBF11\\
         Telco* &  IEEE06, SBF03 & Naval & SBB10\\
         Retail & SBB04, SBF03 & Pharmacy & SPR02\\
         TTT* & SBF03, SBF02 &  & \\
         \bottomrule
    \end{tabular}
    \begin{tablenotes}
       \item  *HRM: Human Resource Management, SM: Sales \& Manufacturing, Tech: Technology/IT, BMC: Business/Management Consulting, Telco: Telecommunication, TTT: Transport,Travel \& Tourism, 
    \end{tablenotes}
    \end{threeparttable}
\end{table}

\begin{table} [htbp]
\centering
\scriptsize
\caption{Categorisation of Studies by SE Phases}
\label{TAB: Categorisation of Studies by SE Phases}
\begin{tabular}{P{0.18\linewidth} P{0.72\linewidth}}
    \toprule
    \textbf{Considered Phase} & \textbf{Study IDs} \\
    \midrule
    
    Planning & IEEE06, SBF04, SBF09 \\
    Requirement Elicitation & IEEE04, SPR01, IEEE06, CHASE01, SBB04, SBB05, SBB06, SBB13, SBF02, SBF04, SPR04, SBF09, SBB17 \\
    Design & IEEE01, SD01, IEEE06, SBF01, SBF04, SBF09 \\
    Implementation & IEEE01, SD01, IEEE06, SBF01, SBF04, SPR03 \\
    Testing & IEEE06, SBF04, SPR03, SBF09 \\
    Maintenance & ACM01, WILEY01, IEEE06, SBB15, SBB14, SBF04, SBF12 \\
    \bottomrule

\end{tabular}
\end{table}

\subsection{Trends, Purpose and Methods of Human Aspects Research (RQ1)}

\subsubsection{Publication Trends}
\begin{figure} [htbp]
    \centering
    \includegraphics[scale = 0.5]{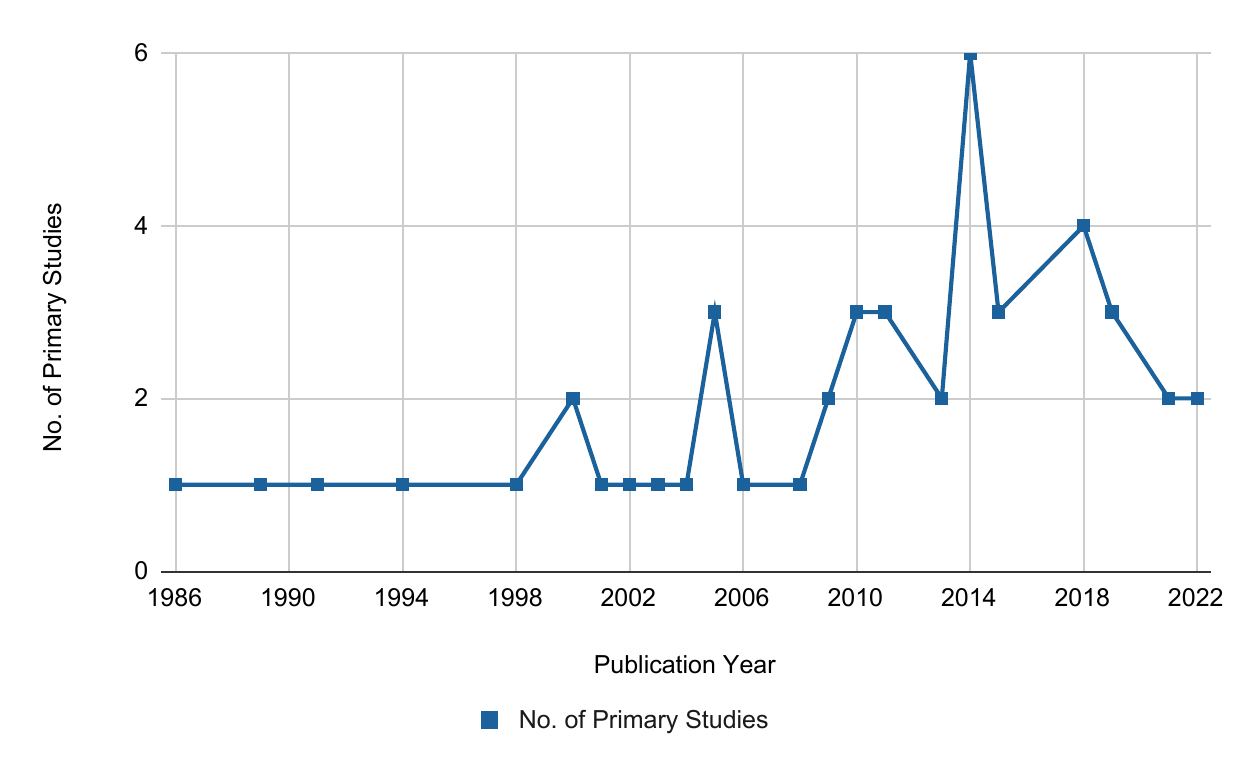}
    \caption{Primary Studies by Year of Publication Year}
    \label{FIG:Primary Studies by Year of Publication Year}
\end{figure}

Figure \ref{FIG:Primary Studies by Year of Publication Year} presents the distribution of primary studies by their respective publication years. The number of papers per year was notably low during the years 1986 to 1998, 2001 to 2004, and 2006 to 2008. However, there were sudden spikes in the years 2000 and 2005. Starting from 2009, the paper count displayed a steady increase, reaching its peak in 2014. However, there were certain years where no primary studies were found.

We analysed the publication trends of human aspects research. Over the years, there has been a consistent effort in academia to explore and understand the dynamics of human aspects between users and developers in SE. 
In the early years of SE research, particularly in the 1980s, there was a notable focus on improving communication between users and developers in the systems development process. Researchers explored methods and strategies to enhance communication channels, understand user requirements, and facilitate collaboration between different stakeholders.
Moving into the 1990s, the focus shifted towards addressing issues related to user involvement and overcoming communication difficulties between users and developers. In this era we observed an increased attention to cultural differences as reasons for communication challenges, with proposed guidelines aimed at fostering mutual understanding between information systems managers and diverse user groups.

In the 2000s, publications focused on overcoming the understanding gap between customers and developers, cultural analysis of communication barriers, and addressing the perception gap between developers and users. There was advocacy for enhancing IT curricula to address the IT-user gap, and research also focused on addressing socio-technical factors causing a developer-user gap. Additionally, research emphasised communication and collaboration between users and developers, and introduced organisational patterns to improve communication in software projects.

The 2010s witnessed a broadening of research topics to include various areas related to human aspects research. There was significant exploration of user involvement, understanding the customer role, managing user-related risks, empathy and usability, emotions, investigating the impact of motivation, and addressing practitioner challenges associated with human factors in software development. Additionally, there was a notable focus on understanding and mitigating developer-user perception gaps, analysing user feedback in app stores, and exploring various aspects of communication and collaboration between users and developers. In 2014, there was a surge in research focusing on the human aspects of developer-user interactions. Studies explored various topics such as understanding communication gaps in large-scale IT projects (SPR02), assessing developers' understanding of usability and empathy (SPR03), investigating the impact of customer interactions on developer motivation (CHASE02), exploring end-user participation methods (SBB07), advocating for better communication between development teams and non-technical customers (SBF06), and presenting empirical findings on usability evaluation methods (SBF08). These studies collectively contribute to a broader understanding of improving communication, collaboration, and empathy between developers and users in software development projects.

\begin{figure} [t]
    \centering
    \includegraphics[width=\linewidth]{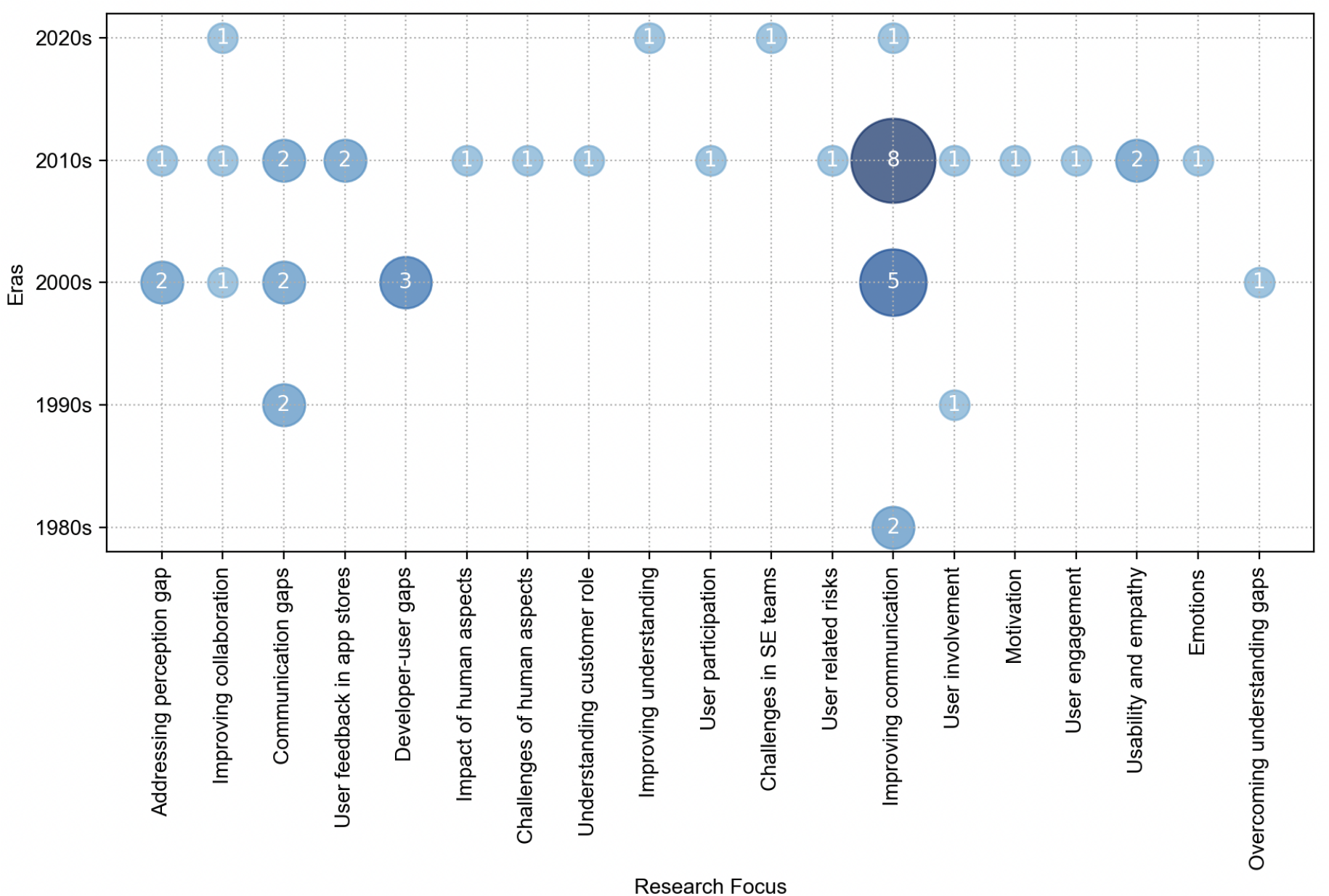}
    \caption{Overview of Publication Trends over the Eras}
    \label{FIG:Publication Trends}
\end{figure}

In recent years (2020s), there has been a continued emphasis on improving communication, understanding, and collaboration among different stakeholders involved in software development. Research has also focused on identifying and addressing challenges faced by SE teams, particularly in areas related to team dynamics, virtualisation, and human factors impacting project success. An overview of publication trends is illustrated in Figure \ref{FIG:Publication Trends}.

\subsubsection{Goals and Objectives of Primary Studies}
Our analysis focused on examining the motivation/goals/objectives of the primary studies. We were keen to understand why these studies delved into the impact of human aspects on the developer-user interactions. We identified four common goals among these studies.

\begin{itemize}
    \item \textbf{Instruments for investigating human aspects}
\end{itemize}

We identified 14 studies that developed new instruments or used existing instruments such as models, tools, frameworks, and guidelines to incorporate human aspects in SE. Some studies adopted a hybrid approach by developing new models based on existing ones, making this the second most common goal.
Among the studies that employed established theories/models, CHASE02 used social-psychological theory to assess motivation in customer-developer interactions, and SBB13 applied the precision model to enhance requirements definition in systems development projects through design team meetings.
There were several studies which developed new models and guidelines such as a questionnaire to measure user engagement success (IEEE02), a set of challenges in SE teams (IEEE03), and a framework to examine the gap between IT professionals and users to identify similarities and differences (SBB08). 
The studies which followed a hybrid approach include, a new variation of cultural probes called infrastructure probes (CHASE01), guidelines for selecting proper communication methods using media richness theory (SBB09), and organisation patterns to improve communication and understanding between development team and customers (SBB10, SBF09). 

SBF01 presented the evaluation of user-developer communication in large-scale IT projects (UDC-LSI) method. The purpose of UDC-LSI method is to increase system success through the increase of UDC in the design and implementation phases. They focused on assessing the system success aspects such as user satisfaction, ease of use etc.
They found an overall positive impact on system success, with the majority of positive responses across all considered system success aspects except for data quality.


SBB19 provided a new perspective into the reasons for the difficulties in communication between users and developers. The authors argued that poor communication may result in failed information systems, excessive maintenance and even increased friction between developers and users. Cultural differences were recognised as the cause of this problem. 
Several guidelines were proposed to assist in developing a mutual understanding of the identified cultural differences.

\begin{itemize}
    \item \textbf{Enhancing the positive impact of human aspects}
\end{itemize}
We identified ten studies that focused on enhancing the positive impact of human aspects resulting in the third most common goal among the included primary studies. In these studies, communication and collaboration aspects facilitated various improvements, such as streamlining usability expertise integration in the software development life cycle (SPR04), enhancing the quality of online user information (IEEE05), and developing better accepted software (SBF04). 
Empathy has helped to get a better understanding of system usability (SPR03) and to improve developer understanding of usability (SBF08). 

Both SPR03 and SBF08 studied empathy. SPR03 explored the origin of novice software developers' empathy towards users and its relation to the improvement in understanding usability. This study found that developer understanding of usability is connected to their empathy towards users. SBF08 investigated the effectiveness of usability evaluations to increase empathy towards users' needs. It showed that the usability evaluation methods that included interaction with users had positive results in generating empathy towards users' needs. Both of these studies found that there is a connection between developer empathy and understanding of usability. 


\begin{itemize}
    \item \textbf{Mitigating the negative impact of human aspects}
\end{itemize}
We found 16 studies focused on mitigating the negative impact of human aspects, making it the most common goal among the included studies. These studies addressed various negative influences, including \textit{perception gaps in requirements understanding}  (SPR01, SBB05, SBB06), \textit{communication gaps} (SPR02, SBF02, SBF05, SBF07, SBB17, SBF10), \textit{collaboration gaps} (SBB07), \textit{developer difficulties in obtaining knowledge about users} (SBB03), and other \textit{challenges related to IT-user interactions} (SBB16, SBF11, SBF06, SBB12).

SBB16 investigated various forms of IT-user gap, such as perspective, ownership, cultural, foresight, communication, expectation, credibility, appreciation, and relationship gaps. The paper proposed solutions, including two curriculum enhancements: one for hybrid majors, stressing both technical and interpersonal skills, and another for hybrid minors, enabling non-IT and non-business majors to develop technical skills.
SBF10 suggested that the difficulties between developers and users are inherent to the cultural differences between these two groups. The study  proposed that a cultural analysis of communication barriers can improve the effectiveness of developer-user interaction. Authors have recommended acknowledging the cultural differences between these groups to improve communication and interaction. 


\begin{itemize}
    \item \textbf{Exploring the nuances of human aspects}
\end{itemize}
We found six studies that explored the nuances of human aspects, making it the least common goal among the included studies. These studies delved into various topics such as \textit{quality requirements in the RE process} (IEEE04), \textit{expressions of emotions and politeness} in developer-user interactions (ACM01), \textit{perceptions of human factors in software development} (WILEY01), \textit{strategies for IT project success} (SBB02), \textit{characteristics of dialogue in user-developer interactions} (SBB14), and \textit{role of the customer in Agile projects} (SBF03).

SBB02 conducted an experiment involving IT project managers, to examine the impact of project partnering, user-developer conflict, and role clarity on software development project performance. They tested a multivariate model and the results supported the model, indicating that user-developer conflict and role ambiguity negatively affect performance estimation difficulty, ultimately impacting project performance negatively.



\subsubsection{Categorisation of Study Methodologies}
\begin{figure}[h!]
    \centering
    \includegraphics[scale=0.5]{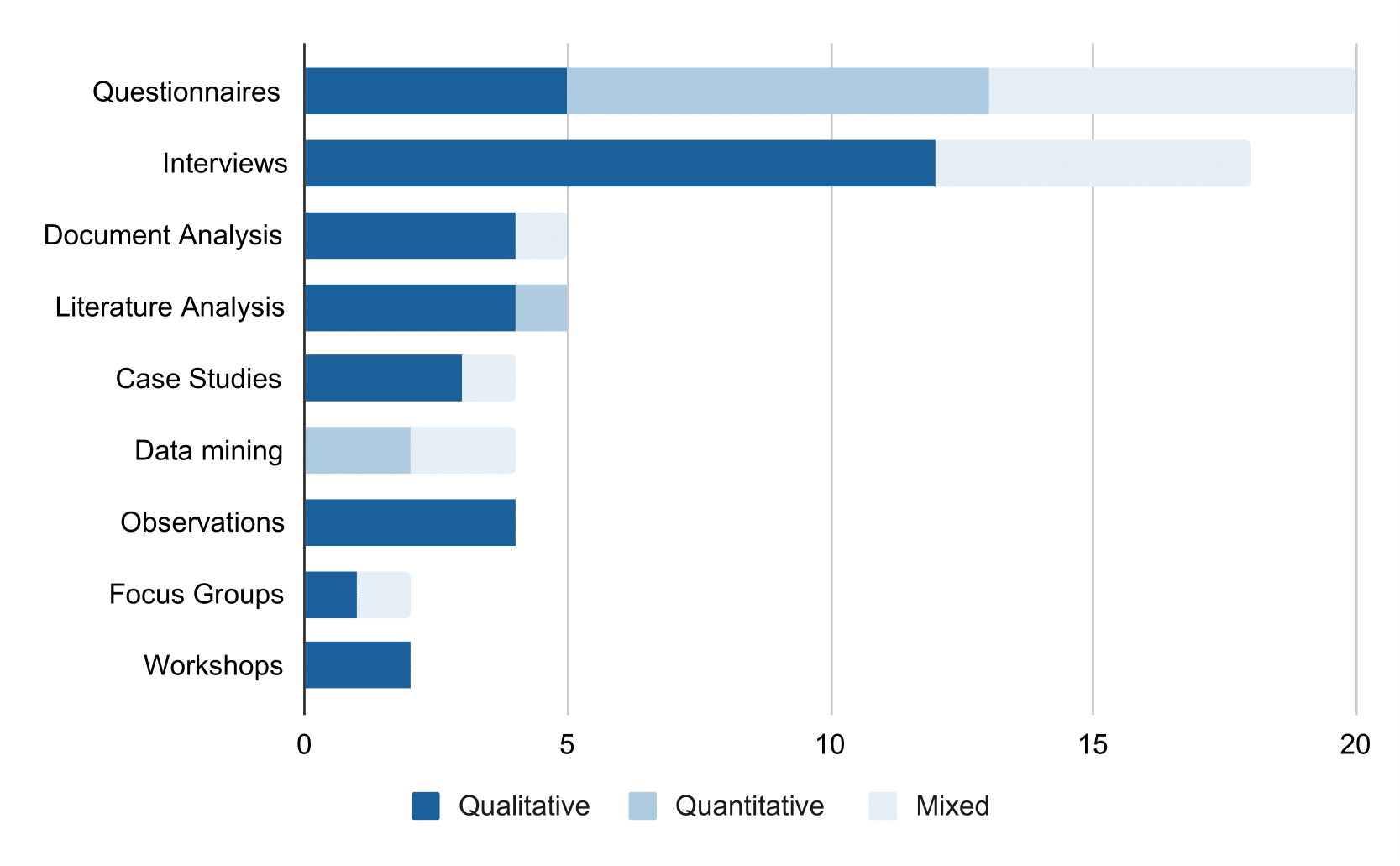}
    \caption{Primary Studies by Data Collection \& Analysis Type}
    \label{FIG:Primary Studies by Data Collection and Analysis Type}
\end{figure}

The primary studies employed various data collection techniques and analysis methods. Questionnaires (21 studies) and interviews (19 studies) were the most commonly used data collection methods. Additionally, observations, document analysis, data mining, case studies, focus groups, workshops, and literature analysis were employed. Most studies (54.35\%) used qualitative data analysis methods, while 21.74\% used quantitative methods, and an equal number used mixed methods. One study did not specify the data analysis type. Figure \ref{FIG:Primary Studies by Data Collection and Analysis Type} illustrates the data collection techniques along with the data analysis methods.

\subsubsection{Categorisation of Study Participants}
The study participants encompassed diverse roles and functions. We categorised them into three main groups: developers, users, and intermediaries who facilitated developer-user interactions. In industry based studies, developers held various roles, such as head of IT department, customer support personnel, project sponsor, technical lead, vendor chief architect, and vendor project manager. In academia based studies, developers mainly consisted of IT undergraduates and postgraduates. Users in industry based studies assumed roles such as internal customers from accounting and marketing departments, user managers, customer product owners, and various other business-related positions. In academia based studies, users included non-IT undergraduates, university department staff, and postgraduates. Intermediaries were human factor experts, usability experts, requirements engineers, IT consultants, and technical communicators/writers. Other participants in various roles included quality assurance staff, agile coaches, designers, senior managers, and more. These participants were engaged in various functions (see Table \ref{TAB:Study Participant Categorisation}).

\begin{table} [t]
    \centering
    \scriptsize
    \caption{Study Participants Categorisation by Functions }
    \label{TAB:Study Participant Categorisation}
    \begin{tabular}{ P{0.3\linewidth} P{0.4\linewidth} P{0.2\linewidth}}
         \toprule
         \textbf{Functions} &  \textbf{Industry} &  \textbf{Academia}\\
         \midrule
         \multicolumn{3}{l}{\textbf{Developer Functions:}}\\
         
         Software designing, development \& defect fixing & IEEE04, SD01, WILEY01, ACM01, SPR02, SBB13, IEEE05, CHASE02, SBB01, SBB03, SBB04, SBB07, SBB08, SBB09, SBB10, SBB15, SBB14, SBF03, SBF01, IEEE03, SBF04, SBF07, SBF05, SBF06, SBF11, SBB12, SPR01, IEEE01 & SBF08, SPR03, IEEE06\\
        
         UI/UX & SPR02, CHASE02 & IEEE06 \\
         Software architectural design & SPR02 & -\\
         Modifying user stories & CHASE02 & - \\
         Customer support activities & SBB07 & -\\
         Conducting evaluations & - & SPR03, IEEE06  \\
         \midrule

         \multicolumn{3}{l}{\textbf{User Functions:}}\\
         Providing domain knowledge & SBB10, IEEE01 & -\\
         Using software & SPR02, IEEE05, CHASE01, SBB01, SBB04, SBB07, SBB08, SBB09, SBB10, SBB11, SBB15, SBB14, SBF01, SBF04, SBF07, SBF05, SBF11, SBF08, IEEE06, SPR01, SD01 & SBB13\\
         Providing feedback & CHASE01, SBB09, SBB11, SBB15, SBB14 & - \\
         Participating in usability evaluations & & SPR03\\
         User testing & SBF03, SPR02 & -\\
         Requirement definition \& verification & CHASE02, SBF03 & -\\
         \midrule

         \multicolumn{3}{l}{\textbf{Functions of Intermediary:}}\\
         Project management activities & - & SBB13\\
         Facilitating communication with users & SPR02, IEEE05, SBB01, SBF07, SD01 & -\\
         Acting as user representatives & SBF01 & -\\
         Usability expertise & SPR04 & -\\
         \bottomrule
    \end{tabular}  
\end{table}

Developers and users employed various interaction methods to engage with each other. The most common method was direct meetings (SD01, SPR02, SBB09-SBB11, SBB17, SBF01, SBF03, SBF05, SBF09), with some conducted online through platforms such as Zoom, Google Meet, and Skype. Additional interaction methods included meetings with customer representatives (SD01, SPR02, SBF03), communication through platforms such as JIRA (SBF01, SBB11), emails (SBB09, SBF05, SBB17), app reviews (SBB15, SBB14, SBF12), telephone calls (SBB09, SBB11, SBB17), and interactions facilitated by intermediaries (SBB01, SBB13). A few studies reported no interaction between developers and users (IEEE04, WILEY01, SBB19). However, many studies did not explicitly mention the interaction methods they used.

\begin{boxA}
    \footnotesize
    \textbf{Answer to RQ1:}
    We categorised primary studies into four common goal categories: instrument development, enhancing positive impact, mitigating negative impact, and exploring nuances of human aspects. The majority of studies focused on mitigating negative impacts. Regarding study methodologies, questionnaires and interviews were frequently used for data collection, and qualitative methods for data analysis. In terms of study participants, we identified three primary groups: developers, users, and intermediaries, each with various roles. Further, we analysed the publication trends of human aspects research over the eras.

\end{boxA}

\subsection{Identified Human Aspects and Study Outcomes (RQ2)}

\subsubsection{Studied Human Aspects}
We identified a range of human aspects that have an impact on the interactions between developers and users. We considered the most commonly described human aspect definitions in the SE context and the definitions used in our primary studies to group the identified human aspects. We categorised these human aspects into three groups: individual, group related, and skill, experiential, or environment-influenced aspects (see Table \ref{TAB:Human Aspects Categorisation}), based on the human aspects taxonomy developed by Grundy et al. \cite{grundy2021addressing}. We defined these three categories of human aspects by extending the original definitions in \cite{grundy2021addressing}. Individual aspects were defined as personal, demographic characteristics that remain relatively stable over the lifetime of a person. Group related aspects are shaped by social contexts and interactions, which may change over time as social dynamics evolve. Skill, experiential, or environment-influenced aspects are context-driven and can change over time due to external factors such as upbringing, training and experience.

\begin{figure} [htbp]
    \centering
    \includegraphics[width=\columnwidth]{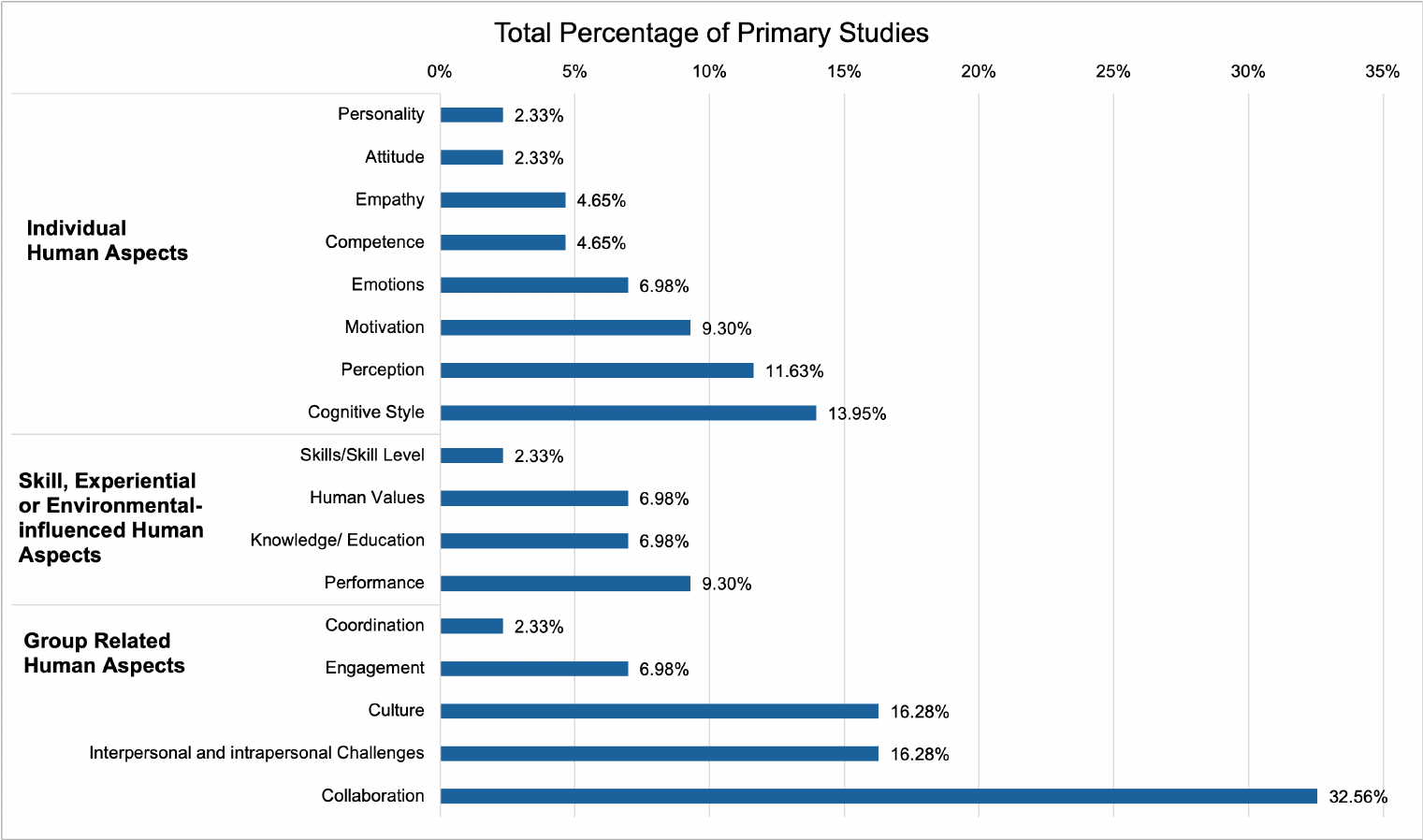}
    \caption{Human Aspects Categorisation}
    \label{FIG:Human Aspects Categorisation}
\end{figure}

The highest number of primary studies were focused on group related and individual human aspects, with only a small number examining skill, experiential, or environment-influenced human aspects. Within group related aspects, communication was the most explored (62.79\%), followed by collaboration (32.56\%), culture (16.28\%), interpersonal and intrapersonal challenges (16.28\%), engagement (6.98\%), and coordination (2.33\%). Among individual human aspects, perception and cognitive style were the most researched (11.63\% each), followed by motivation (9.3\%), emotions (6.98\%), empathy (4.65\%), competence (4.65\%), personality (2.33\%), and attitude (2.33\%). In the skill, experiential, or environment-influenced aspects, the majority of studies were related to performance (9.30\%), while others addressed human values (6.98\%), knowledge/education (6.98\%), and skills/skill level (2.33\%). 
When considering all the human aspects, skills, personality, coordination, attitude, empathy, and competence were the least studied human aspects in the primary studies as illustrated in Figure \ref{FIG:Human Aspects Categorisation}.

\begin{table} [ht]
    \centering
    \begin{threeparttable}
    \scriptsize
    \caption{Human Aspects Categorisation}
    \label{TAB:Human Aspects Categorisation}
    \begin{tabular}{P{0.05\linewidth} P{0.22\linewidth} P{0.65\linewidth}}
         \toprule
         \textbf{Category} & \textbf{Human Aspect} & \textbf{Study IDs}\\
         \midrule
           
         \multirow{8}{*}{\rotatebox[origin=c]{90} {\parbox[c]{2.5cm}{\centering Individual Human Aspects}}} & Personality & SBB12 \\
         & Attitude & IEEE02\\
         & Empathy & SPR03, SBF08\\
         & Competence & IEEE02, SBB01\\
         & Emotions & ACM01, IEEE04, SBB16 \\
         & Motivation & IEEE02, CHASE02, SBB01, SBB12  \\
         & Perception & SPR01, SBB05, SBB06, SBB16, SBF06\\
         & Cognitive Style & IEEE01, IEEE04, WILEY01, SBB04, SBB16, SBF09\\
         \midrule
         
         \multirow{4}{*}{\rotatebox[origin=c]{90} {\parbox[c]{1.4cm}{\centering Skill,Exp/ Env-infl*}}} & Skills/Skill Level & WILEY01 \\
         & Human Values & IEEE04, ACM01, SBB12\\
         & Knowledge/Education & IEEE04, SBB04, WILEY01\\
         & Performance & SBB01, SBB02, SBB05, SBB06\\
         \midrule
         
        \multirow{6}{*}{\rotatebox[origin=c]{90} {\parbox[c]{3.5cm}{\centering Group Related Human Aspects}}} & Coordination & SBB05 \\
         & Engagement & IEEE02, SBB01, WILEY01\\
         & Culture & IEEE02, SBB04, SBB12, SBB16, SBB19, SBF07, SBF10 \\
         & Interpersonal \& Intrapersonal Challenges & IEEE02, IEEE03, SBB02, SBB03, SBB08, SBB16, SBF11\\
         & Collaboration & SBB12, IEEE06, SBB01, SBB07, SBB10, SBF03, SBF04, SBF05, SBB05, SBB02, SBB16, IEEE02, WILEY01, SD01\\
         & Communication & IEEE01, IEEE02, IEEE04, ACM01, SD01, SPR02, SBB12, IEEE05, CHASE01, SBB05, SBB09, SBB10, SBB11, SBB13, SBB19, SBB15, SBB16, SBB14, SBB18, SBF01, SBF02, SBF04, SBF07, SBF05, SBF10, SBB06, SPR04, SBF09, SBB17, SBF12\\
        \bottomrule
    \end{tabular}
    \begin{tablenotes}
        \item Skill, Exp/ Env-infl*: Skill, Experiential or Environmental-influenced Human Aspects
    \end{tablenotes}
    \end{threeparttable}
\end{table}

\subsubsection{Relationships between Human Aspects}
Among the primary studies, only about 14\% explored the relationships between different human aspects. These studies revealed both positive and negative relationships (see Table \ref{TAB:Categorisation by Relationships between Human Aspects}). The most discussed relationships included positive connections between culture and communication and performance and collaboration. Cultural differences were identified as a foundational factor for communication gaps, and organisational culture influenced user-developer communication strategies (SBB19, SBF07, SBF10). Collaboration between developers and users was found to enhance performance (SBB01, SBB02). Negative relationships were observed between interpersonal and intrapersonal challenges and performance, as well as collaboration and these challenges. In both cases, these challenges had adverse effects on the other human aspects (SBB02). Another interesting finding was the positive relationship between culture and these challenges. While individuals' values didn't significantly influence interpersonal challenges, cultural differences led to more frequent and critical challenges (IEEE03).

\begin{table} [ht]
\centering
\begin{threeparttable}
    \tiny
    \caption{Categorisation by Relationships between Human Aspects}
    \label{TAB:Categorisation by Relationships between Human Aspects}
    \begin{tabular}{P{0.05\linewidth} P{0.15\linewidth} P{0.7\linewidth}}
        \toprule
        \textbf{Type} & \textbf{Human Aspects} & \textbf{Paper IDs and Relationship Details} \\
        \midrule
    
        \multirow{9}{*}{\rotatebox[origin=c]{90} {\parbox[c]{4cm}{\centering Positive Relationships}}} & Competence, Engagement & \textbf{SBB01}: User ability (competence) is positively associated with user commitment (engagement) to the IS project team. \\ 
        
        & Motivation, Engagement & \textbf{SBB01}: Extrinsic motivation is positively associated with user commitment to the IS development project.  \\
        
        & Competence, Collaboration & \textbf{SBB01}: Ability and extrinsic motivation are positively and strongly correlated with user–IS collaboration.\\ 
        
        & Motivation, Collaboration & \textbf{SBB01}: Ability and extrinsic motivation are positively and strongly correlated with user–IS collaboration.\\
        
        & Performance, Collaboration & \textbf{SBB01}: User–IS collaboration is positively associated with project performance.  
        
        \textbf{SBB02}: Partnering is negatively associated with performance estimation difficulty.
        Partnering is positively associated with project performance.
        Performance estimation difficulty is negatively associated with project performance. \\
        
        & Engagement, Collaboration & \textbf{SBB01}: User commitment is positively associated with user–IS collaboration. \\
        
        & Engagement, Performance & \textbf{SBB01}: User commitment to the IS development team is positively associated with project performance. \\
        
        & Communication, Culture & \textbf{SBB19}: Cultural differences between users and developers are a major cause of communication gaps. 
        
        \textbf{SBF07}: User-developer communication strategies can be seen as a function of organisational culture. They proposed to use a technical writer to act as a mediator of communication between stakeholders. 
        
        \textbf{SBF10}: Communication problems can be explained in term of inherent cultural differences between departmental communities and an understanding and appreciation of cultural behaviour can help improve interdepartmental relations. Differences in national culture also have an impact on communication between two groups.\\ 
        
        & Culture, Challenges* & \textbf{IEEE03}: Teams with 2-3 nationalities among the team members had less frequent and critical challenges than teams with just a single nation.\\ 
        \midrule
        
        \multirow{2}{*}{\rotatebox[origin=c]{90} {\parbox[c]{1.2cm}{\centering Negative Relationships}}} & Challenges*, Performance & \textbf{SBB02}: User–developer conflict (challenges) is negatively associated with project performance.
        User–developer conflict is positively associated with performance estimation difficulty. \\
        
        & Collaboration, Challenges* & \textbf{SBB02}: Partnering is negatively associated with user–developer conflict. \\
        \bottomrule
    
    \end{tabular}
        \begin{tablenotes}
            \item *Challenges: Interpersonal and Intrapersonal Challenges
        \end{tablenotes}
\end{threeparttable}
\end{table}

\subsubsection{Categorisation of Study Outcomes}
The outcomes or solutions provided by the studies fell into four main categories: working tools, models, prototypes, or facilitating tool development; theories, theoretical frameworks, or theoretical models; guidelines, approaches, methods, or practices; and empirical insights, indications, or evidence on human aspects in developer-user interactions (see Table \ref{TAB:Study Categorisation by Outcome}).

\begin{table}[htbp]
\tiny
\centering
\caption{Study Categorisation by Outcomes}
\label{TAB:Study Categorisation by Outcome}
\begin{tabular}{P{0.07\linewidth} P{0.1\linewidth} P{0.73\linewidth}}
    \toprule
    \textbf{Category} & \textbf{Paper ID} & \textbf{Study Outcome} \\
    \midrule
    
    \multirow{10}{*}{\rotatebox[origin=c]{90} {\parbox[c]{3cm}{\centering Working tool/ Model/ Prototype or Facilitating tool development}}} & IEEE01 & A Descriptive classification for end user-relevant decisions of large-scale IT projects \\
    
     & IEEE02 & A questionnaire to measure user engagement success \\
     & CHASE01 & Infrastructure probes (IP) \\
     & SBB08 & Gap model \\
     & SBB11 & A process model for Communication \\
     & SBB13 & A meeting format using the set of communication behaviours contained in the precision model  \\
     & SBB18 & An integrated framework of communication \\
     & SBF01 & Evaluation of the UCD-LSI method \\
     & SBF04 & Facilitating the development of CloudTeams platform \\
     & SPR04 & RESPECT framework, framework for user-centred and use-case driven RE, XML-based tool for identifying, studying and mediating human-to-human communication \\
    \midrule
    
    \multirow{9}{*}{\rotatebox[origin=c]{90} {\parbox[c]{4cm}{\centering Theory/ Theoretical Framework/ Theoretical Model}}} & IEEE04 & Five data patterns related to delivery of quality requirements and how it affects project success \\
     & SPR01 & Five hypotheses related to: user participation and perception gap, requirements uncertainty and perception gap, requirements uncertainty and user participation, top management support, perception gap, and user participation \\
     & IEEE06 & Hypotheses were drawn based on the customer-driven courses \\
     & SBB01 & Hypotheses were drawn based on: user ability and user commitment, extrinsic motivation and user commitment, user commitment and IS–user collaboration, user commitment and project performance, user–IS team collaboration and project performance. \\
     & SBB02 & Hypotheses were drawn based on relation of project partnering, user–developer conflict, and lack of role clarity to the performance of software development projects \\
     & SBB05 & Hypotheses were drawn based on pre-project partnering, perception gap, horizontal coordination, vertical coordination and project performance \\
     & SBB06 & Hypotheses were drawn based on requirements instability, stakeholder perception gap, requirements diversity, residual performance risk and project management performance \\
     & SBB10 & Four organisation patterns \\
     & SBF09 & Six organisational patterns \\
     \midrule

    \multirow{10}{*}{\rotatebox[origin=c]{90} {\parbox[c]{3.5cm}{\centering Guidelines/ Approach/ Methods/ Practices}}} & SPR02 & Ideas to overcome obstacles for the implementation of UDC and factors for communication gaps \\
     & SBB03 & Approach for overcoming the user involvement obstacles \\
     & SBB04 & An approach to help overcome some of the mismatch or understanding gap between the customer and the developer \\
     & SBB19 & Guidelines for bridging cultural gap  \\
     & SBB16 & Enhancements to the traditional IT curriculum \\
     & SBF03 & Practices to enhance the effectiveness of the on-site customer \\
     & SBF06 & List of methods to improve knowledge of the non-technical clients in order to bridge the gap between non-technical clients and technical professionals \\
     & SBF10 & Recommendations to improve communication in information systems development  \\
     & SBF11 & Solutions to bridge the gaps between users and developers \\
     & SBB17 & Intervention for Managing the Challenges \\
    \midrule
    
    \multirow{17}{*}{\rotatebox[origin=c]{90} {\parbox[c]{4cm}{\centering Empirical insights/ Indications/ Evidence on human aspects in developer-user interactions}}} & ACM01 & Insights on emotions, politeness, communication  \\
     & WILEY01 & Insights on employee awareness, leadership involvement, employee involvement, customer involvement, senior management support, staff experience, staff learning, staff skills, client support, software process improvement (SPI) consultancy, reward schemes \\
     & SD01 & Insights on communication, collaboration \\
     & SBB12 & Insights on collaboration, organisational culture, trust, cohesion, personality, motivation, communication  \\
     & IEEE05 & Insights on communication  \\
     & CHASE02 & Insights on motivation  \\
     & SBB07 & Insights on collaboration  \\
     & SBB09 & Insights on communication  \\
     & SBB15 & Insights on communication  \\
     & SBB14 & Insights on communication  \\
     & IEEE03 & Insights on interpersonal and intrapersonal challenges  \\
     & SBF02 & Insights on communication  \\
     & SBF07 & Insights on communication, culture gap  \\
     & SPR03 & Insights on empathy  \\
     & SBF05 & Insights on communication, collaboration  \\
     & SBF08 & Insights on empathy  \\
     & SBF12 & Insights on communication via app reviews \\
     \bottomrule

\end{tabular}
\end{table}

\begin{itemize}
    \item \textbf{Working Tool/Model/Prototype or Facilitating Tool Development}
\end{itemize}
The second highest number of studies provided working tools, models, prototypes, or facilitated tool development and evaluation related to human aspects. For example, SPR04 proposed the RESPECT framework for user-centred and use-case driven RE. RESPECT contains a set of principles to improve and mediate software-to-usability communication involved in identifying complementarities between the use-case requirements and RESPECT processes. The XML-Based tool for identifying, studying and mediating human-to-human communication known as SUCRE Mediator, aids in studying and comparing processes and mediating communication between usability and software engineers. 
CHASE01 presented infrastructure probes (IP), a new variation of cultural probes. This can be seen as an ethnographic method to get a deeper understanding of the user’s working context thus help to improve the collaboration between developers and users regarding requirements elicitation. SBB13 presented a meeting format using a set of communication behaviours contained in the Precision Model. This format can be incorporated into a general format for running team meetings and interviews, helping developers to better fill in information gaps in users’ specifications and generating higher quality information during requirements definition.

\begin{itemize}
    \item \textbf{Theory/Theoretical Framework/Theoretical Model}
\end{itemize}
A few studies produced a theory, theoretical framework, or theoretical model as an outcome, with this category having the lowest number of studies. For example, IEEE04 examined how human values, particularly trustworthiness, contribute to successful communication between developers and users. This study explored the factors influencing the delivery of quality requirements and their impact on project success by presenting data patterns. SBB10 found four organisational patterns that complement agile methods by establishing a better baseline of requirements and a better customer-developer working relationship. It was shown that this relationship is strengthened throughout the project life cycle by continually applying these patterns. In SBF09, researchers found that lack of customer involvement, absence of a common communication language, complex business processes, end-user resistance toward system development, lack of trust, lack of leadership, and end-user capacity as challenges that hinder the understanding of the development team.
In addition to theoretical frameworks and models, this category includes various hypotheses derived from different factors, such as customer-driven courses, user ability, user commitment, extrinsic motivation, IS–user collaboration, and project performance (see Table \ref{TAB:Study Categorisation by Outcome}).

\begin{itemize}
    \item \textbf{Guidelines/Approaches/Methods/ Practices}
\end{itemize}
The third-highest number of studies provided lists of guidelines, approaches, or methods aimed at improving SE outcomes by integrating human aspects or mitigating the negative effects caused by a lack of consideration of human aspects. For instance, SBB04 developed an approach to address understanding gaps between customers and developers. The study emphasised the importance of capturing accurate user requirements early in the development process, attributing challenges to cultural and knowledge gaps between users and developers. To bridge these gaps, the study introduced the use of various set theory diagrams to establish shared understanding between users and developers. SBF06 discussed perceptions of software development practitioners on competence of their customers. The authors argued that one potential way to solve the problem is establishing a common language between development team and non-technical customers, and they proposed the first step of establishing such a common language for information systems development. SBB17 developed an intervention for managing the communication challenges faced by developers and customers during  requirements elicitation. The challenges include, understanding real customer requirements, ambiguities in the expression of customer requirements, medium of communication, developers' knowledge and developer ability to express requirements.

\begin{itemize}
    \item \textbf{Empirical Insights/Indications/Evidence on Human Aspects in Developer-User Interactions}
\end{itemize}
The majority of the studies offered empirical insights, indications, or evidence related to human aspects as their outcomes.
For instance, SD01 provided insights into communication. The study presented a case study involving the collaboration of human factor (HF) experts, users, and a company developing medical software. Their research highlighted the benefits of integrating user representatives into the software life cycle for end-users. However, it also pointed out that this approach alone might not be sufficient to resolve complex usability issues. 
Similarly, IEEE05 shared insights on communication. This paper discussed how involving technical communicators in writing online help and crafting system and error messages improved customer satisfaction. The study emphasised that the work of technical communicators added value to the development process by enhancing the quality of information products in these areas. 
SBB07 provided insights on collaboration. This is a study of a health information system development, that analysed the opinions of physicians (end-users) and developers regarding end-user participation in application development. 
IEEE03 provided insights into interpersonal and intrapersonal challenges. This study presented a set of relevant human challenges in SE teams, and a refined set of challenges between colleagues (interpersonal), and challenges between an individual and their work (intrapersonal).

\subsubsection{Evaluation Methods Used}
The evaluation methods used in the primary studies varied, with only 19 out of the total 46 studies having evaluated their developed solutions (see Table \ref{TAB:Study Categorisation by Evaluation Methods}). The most commonly adopted evaluation methods included statistical analysis (5 studies) and validation using an existing model, theory, or framework (5 studies). Other frequently used evaluation methods were expert feedback (3 studies) and application in real software development projects (3 studies). Only a few studies employed prototypes (1 study), case studies (1 study), and empirical testing methods such as structural equation modelling (1 study). It is worth noting that 27 studies did not evaluate their study outcomes and we identified two main reasons for this. Firstly, in some cases, the evaluation process had not been completed at the time of publication, and was outlined as part of the authors' future work. Secondly, there were instances where evaluation details were neither provided in the paper nor mentioned as future work. In some of these cases, the specific reason for the omission was unclear.
However, it is important to note that not all studies may necessitate formal evaluation processes. For instance, empirical studies focusing on industry practices or qualitative research involving interviews may primarily report on findings derived from real-world data rather than evaluating specific interventions or methodologies \cite{busetto2020qualitative}. This aligns with the notion that empirical work often involves capturing and analysing existing phenomena or real-world data rather than assessing the effectiveness of interventions. We identified several studies which may not require a formal evaluation due to the empirical nature of their studies (IEEE03, IEEE04, IEEE05, IEEE06, SPR02, SBB03, SBB08, SBB11, SBB12, SBB17, SBF02, SBF03, SBF04, SBF05, SBF06, SBF07, SBF09, SBF11). While the absence of explicit evaluation details may not always imply unreliability, clear and comprehensive reporting is essential for upholding the integrity of research outcomes. It serves to verify the adherence to methodological criteria, thereby ensuring the reliability of the findings.

\begin{table}[htbp]
\scriptsize
\centering
\caption{Study Categorisation by Evaluation Methods}
\label{TAB:Study Categorisation by Evaluation Methods}
\begin{tabular}{P{0.5\linewidth} P{0.43\linewidth}}
    \toprule
    \textbf{Evaluation Method} & \textbf{Study IDs} \\
    \midrule
    
    Expert feedback & IEEE01, IEEE02, WILEY01 \\
    Using a prototype & CHASE01 \\
    Using statistical analysis (descriptive statistics) & SBB01, SBB05, SBB06, SBB02, SBB15 \\
    Via a case study & SBF01 \\
    Applying or complementing an existing model/ theory/framework & SPR03, CHASE02, SBB04, SBB09, SBF08 \\
    Via empirical testing methods & SPR01 \\
    Applying in practice (in real software development) & SBB10, SBB13, SPR04 \\
    \bottomrule
    
\end{tabular}
\end{table}

 \begin{boxA}
    \footnotesize
    \textbf{Answer to RQ2:}
    We categorised various human aspects into individual, group-related, and skill, experiential, or environment-influenced categories, impacting developer-user interactions. Most primary studies focused on group-related aspects, with communication being the most studied. We examined both positive and negative relationships among these aspects, with only around 14\% identifying such connections. Study outcomes fell into four main categories: tool development, theoretical frameworks, guidelines, and empirical insights on human aspects. The majority of studies provided empirical insights as their outcome. Most studies used statistical analysis and validation through existing models or theories for evaluation. However, the majority of studies did not report details regarding the evaluation. We also identified several studies which may not require a formal evaluation due to the empirical nature of their studies.  
\end{boxA}

\subsection{Nature of the Impact of Human Aspects (RQ3)}

\begin{figure}[htbp]
    \centering
    \includegraphics[scale=0.4]{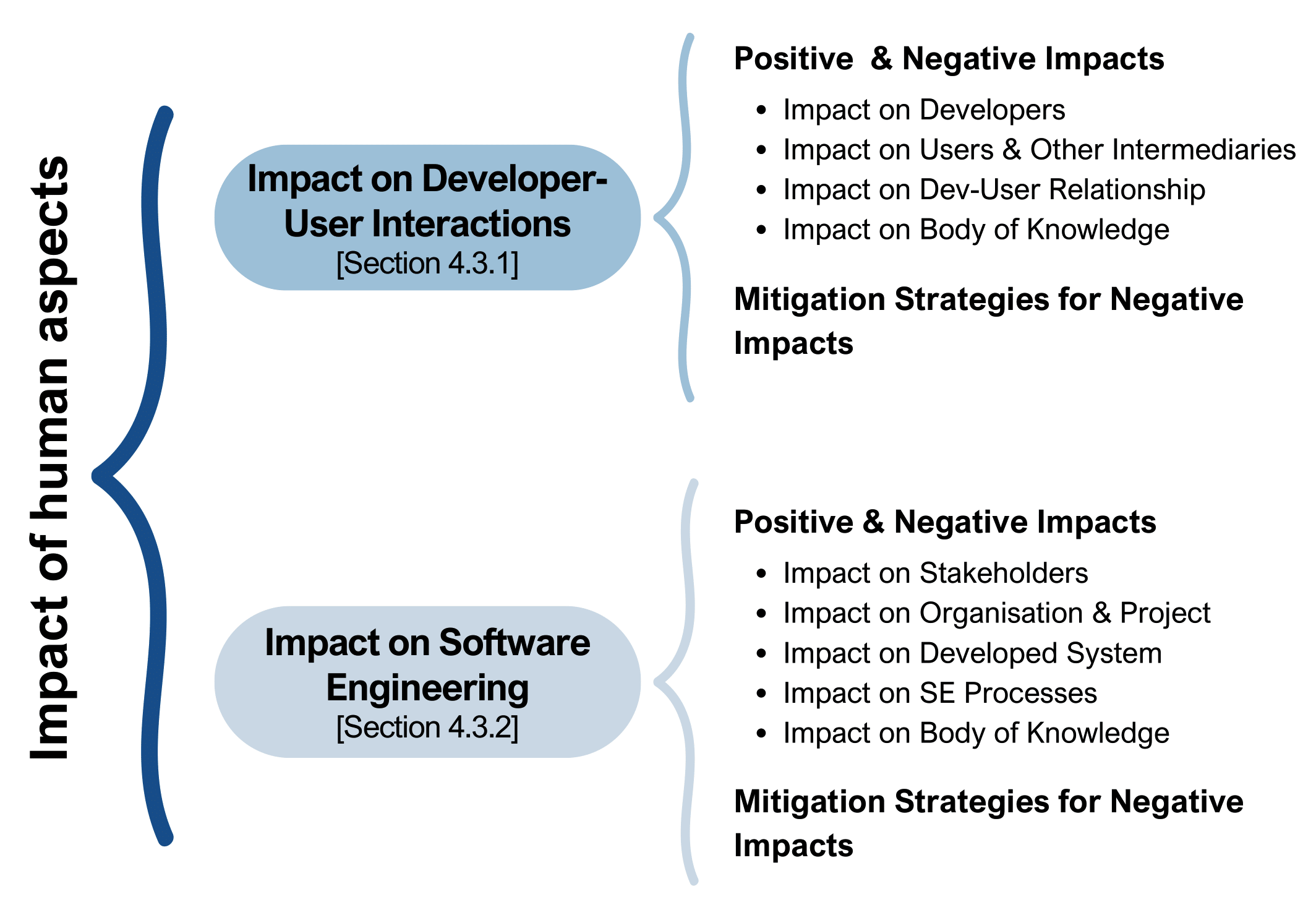}
    \caption{Impact of Human Aspects Categorisation (Overview of Section 4.3)}
    \label{FIG:Impact of Human Aspects Categorisation}
\end{figure}

In all the primary studies, the nature of the impact of human aspects in developer-user interactions was identified and categorised as either positive, negative, or mixed. The majority of the studies (27 studies) identified positive effects of human aspects, while 12 studies found negative effects. Seven studies identified both positive and negative effects. We categorised these effects as shown in Figure \ref{FIG:Impact of Human Aspects Categorisation}. 
Some of the positive and negative effects are summarised in \ref{Appendix:Positive Effects of Human Aspects} and \ref{Appendix:Negative Effects of Human Aspects}. For a more comprehensive analysis of both positive and negative effects of human aspects, the supplementary information package is accessible online. \footnote{https://github.com/Hashini-G/SupplementaryInfoPackage-SLR}
 
\subsubsection{Impact on Developer-User Interactions}

We categorised positive and negative effects on developer-user interactions into four categories as effects on: developers, users and other intermediaries, the relationship between developers and users, and the body of knowledge. 

\begin{itemize}
    \item \textbf{Impact on Developers}
\end{itemize}
This category encompasses positive and negative impacts attributed to developers or those caused by developers' behaviours. In this category, we observed that developers experienced positive impacts such as increased empathy towards users and their needs, heightened motivation, improved technical, soft, and project management skills, enhanced work efficiency, greater buy-in and ownership, improved productivity, increased appreciation for customer tasks, and a sense of empowerment and motivation from workload control. These effects were attributed to various human aspects, including empathy, motivation, collaboration, and communication. 
For example, in SBF03, the study explored how the customer role can be effectively implemented in Agile projects. This study found that both developers and customers enjoyed direct interactions, which helped developers better understand the extent of the customer tasks and increased their appreciation for these tasks.

Negative effects comprise developers' unawareness of customer needs, limited domain knowledge, demotivation among developers, negative developer attitudes towards users, increased pressure and stress on developers, and developers being misled by users. These impacts predominantly stemmed from communication and interpersonal and intrapersonal challenges.

\begin{itemize}
    \item \textbf{Impact on Users and Other Intermediaries }
\end{itemize}
This category encompasses positive and negative impacts on users and intermediaries, or those caused by their behaviours. We identified positive impacts such as improved client collaboration with developers, increased customer satisfaction, enhanced learning opportunities, greater participation, positive user attitudes, effective system use, and improved information flow. These effects were the result of various human aspects, including perception, emotions, communication, and engagement. For example, SBF06 discussed the perceptions of software developers regarding the competence of their customers. They recommended fostering a common language and common expectations between development team and non-technical clients. This initiative was found to enhance client collaboration with developers.

The negative effects include customers feeling a lack of control over scope and schedule, decreased user involvement, negative user attitudes toward technology, and unrealistic user expectations resulting in stress and dissatisfaction with developers. These issues are primarily driven by human aspects including motivation, perception, performance, communication, collaboration, coordination, and interpersonal \& intrapersonal challenges.

\begin{itemize}
    \item \textbf{Impact on the Relationship between Developers \& Users}
\end{itemize}
This category covers positive and negative effects on the developer-user relationship or those resulting from this relationship. We identified various positive effects, including a reduced developer-user perception gap, development and maintenance of a strong developer-user rapport, improved understanding, increased trust, enhanced knowledge sharing, enabling richer communication, improved collaboration, reduced conflicts, enhanced coordination, and increased awareness of developer-user differences. These effects were influenced by several human aspects, such as perception, emotions, communication, collaboration, cognitive style, culture, interpersonal and intrapersonal challenges, human values, knowledge, coordination, and engagement. For instance, in SBB05, the study explored developer-user perception gaps and their adverse effects on project performance. The study uncovered that pre-project partnering substantially enhances coordination and reduces these perception gaps, resulting in a more positive developer-user relationship.

These negative impacts encompass a lack of trust, reduced collaboration, conflicts, communication breakdown, increased perception gaps, lack of understanding among stakeholders, strained relationships, resistance between developers and users, cultural gap, misaligned stakeholder objectives, and a lack of empathy. These issues are influenced by human aspects such as motivation, emotions, cognitive style, human values, knowledge, communication, interpersonal and intrapersonal challenges, perception, performance, collaboration, coordination, and culture.

\begin{itemize}
    \item \textbf{Impact on the Body of Knowledge}
\end{itemize}
We placed impact on the body of knowledge into this category, and only the positive effects were found. These effects included refined and improved system requirements, valuable insights regarding developer-user interaction, and increased awareness of user participation and collaboration. Multiple human factors, such as knowledge, culture, communication, and collaboration, contributed to these effects. For example, SBB07 is a study centred on a health information system. It discussed the opinions of physicians (users) and developers regarding end-user participation in application development. This study found that both users and developers are willing to collaborate, leading to valuable insights about their interactions and contributing to the overall body of knowledge in this domain.

\begin{itemize}
    \item \textbf{Mitigation Strategies for Negative Impact on Developer-User Interactions}
\end{itemize}
We identified various strategies employed by primary studies to mitigate the negative impact on developer-user interactions. Managerial interventions such as coordination and partnering helped to reduce perception gaps and enhance project performance. Building collaborative relationships among all stakeholders fostered cooperation and understanding. Developing hybrid positions, such as business analysts, technical writers helped bridge the IT-user gap, facilitating communication between technical and non-technical areas. Also employing existing team members as communication facilitators and involving business analysts or technical writers to bridge developer-user gap facilitated mutual understanding. Continuous interaction with users, maintaining open communication, and avoiding specialist terminology ensured clarity and alignment in information shared between them. We also found that, it is important to acknowledge that the computer-based systems serve only as a tool to support work for business users, whereas for developers, system development is their primary responsibility. Treating business users as unique individuals and appreciating their specific contributions helped to enhance development. Further, considering business users as clients and dedicating time to understand their needs was considered vital for improving software development. Collaborative user partnering reduced conflicts while enhancing interaction and communication with IS staff. 
Further we found some strategies that can be implemented via project managers. Project managers should promote an environment where users feel empowered to share their ideas openly. Creating a `safe space' for both users and developers to be creative is essential, possibly facilitated by external facilitators or internal experienced individuals. Avoiding reliance on a limited group of users, implementing a cautious selection process for appointing users to the design team, employing specialised communication methods (e.g. precision model) or supplementary techniques such as contextual inquiry, which permit designers to spend time observing users in their native environment are recommended during design meetings.
The Gap model, which highlights differences between users and developers, can be used to understand real world disparities. These strategies will help to navigate the complexities of human aspects in developer-user interactions effectively.


\subsubsection{Impact on SE}
We identified five groups of positive and negative effects on SE, including impact on: stakeholders and their relationships, the organisation and project, the developed systems, SE processes, and body of knowledge. 

\begin{itemize}
    \item \textbf{Impact on Stakeholders/ Stakeholder Relationships}
\end{itemize}
This category encompasses the positive and negative impact on developers, users, and other stakeholders, along with their relationships. 
Positive impacts include enhanced team dynamics (e.g., increased team motivation, involvement, customer and developer satisfaction), better understanding of the problem domain, improved understanding of the developers' thought process, assistance in decision making, enhanced user engagement \& participation, and encouraging users to reflect on their technology use. Additionally, it fostered innovation, improved partnerships in projects, improved task management abilities of developers, assisted in overcoming team limitations, promoted ongoing cooperation between customers and developers, facilitated the acquisition of in-depth domain knowledge, and assisted in mitigating conflicts between developers and users.
These effects are influenced by human aspects such as empathy, communication, collaboration, culture, engagement, emotions, cognitive style, and knowledge.

Negative effects include reduced collaboration between developers and users, decreased user participation, diminished performance, a lack of shared understanding, incomplete domain knowledge, lower customer satisfaction, limited customer involvement, reduced developer-user communication, a lack of trust, sparse user feedback, decreased team cohesion, subjective task interpretations, inadequate initial task analysis, slow decision-making, imbalance between responsibility and authority, resistance to suggestions, and customer resistance to software projects. These effects are influenced by human aspects including motivation, performance, collaboration, interpersonal and intrapersonal challenges, perception, communication, coordination, emotions, cognitive style, human values, and knowledge.

\begin{itemize}
    \item \textbf{Impact on Organisation \& Software Project}
\end{itemize}
This category involves positive and negative impacts on software project, software-producing organisation, and other resources and infrastructure of the software project, including budget and time. Positive effects involve resource optimisation, enhanced project success, increased productivity, a more collaborative organisational culture, improved project performance, assistance in release planning, boosted project momentum, and support for various project phases and activities. These outcomes are influenced by human aspects including motivation, emotions, communication, cognitive style, performance, collaboration, coordination, engagement, and empathy.

Negative effects comprise a lack of top management support for user participation, difficulties in estimating project schedule and cost, poorly coordinated development efforts, increased project failures, greater challenges in estimating project performance, hindrances to effective business practices, incurring costs for the software development organisation, obstacles in system design and development, incorrect allocation of time, staff, and resources, a lack of necessary expertise in the development team, and insufficient project documentation. These effects are influenced by human aspects including perception, performance, communication, emotions, cognitive style, collaboration, culture, interpersonal and intrapersonal challenges, human values, knowledge, and coordination.

\begin{itemize}
    \item \textbf{Impact on the Developed System}
\end{itemize}
This category covers the positive and negative impact on the developed software system and its consequences. Positive impacts encompass enhanced system usability, increased system success, generation of refined and improved system requirements leading to systems that better meet stakeholders' requirements and expectations, improved software quality, enhanced data quality, improved user experience, increased system usage, reduced defect rates, and an improved user support process. These improvements are influenced by human aspects such as empathy, communication, collaboration, culture, engagement, emotions, cognitive style, and knowledge/education.

Negative effects involve issues such as requirements uncertainty, diverse requirements, unrealistic specifications that fail to meet user needs, increased system failures, information systems' inability to meet business needs, reduced usage of information systems, excessive maintenance to fulfil requirements, users feeling threatened by the systems, and a lack of confidence in the system. These outcomes are influenced by human aspects including perception, emotions, cognitive style, human values, knowledge, performance, communication, collaboration, culture, coordination, and interpersonal and intrapersonal challenges.

\begin{itemize}
    \item \textbf{Impact on SE Processes }
\end{itemize}
This category involves positive and negative impacts on various SE-related activities, processes, and phases. Positive effects include improved problem-solving for complex SE issues, support for successful implementation of software process improvement(SPI) efforts, and fostering the formation of startups. These positive effects are influenced by human aspects including personality, communication, culture, cognitive style, knowledge, skills, collaboration, and engagement.

Negative effects include unsuccessful (SPI) efforts in SME software development companies and challenges in integrating usability expertise into the software development life cycle. These impacts are attributed to human aspects including cognitive style, knowledge, skills, collaboration, engagement, and communication.

\begin{itemize}
    \item \textbf{Impact on the Body of Knowledge}
\end{itemize}
This category covers the impact on the existing body of knowledge, which involves contributing to the creation of new knowledge or improving and strengthening existing knowledge. We identified only the positive effects under this category. These effects encompass an enhanced understanding of usability, insights into the impact of human aspects, and contributions to the research of human aspects in SE. These outcomes are influenced by factors including empathy, motivation, attitude, competence, communication, collaboration, culture, interpersonal and intrapersonal challenges, emotions, and human values.

\begin{itemize}
    \item \textbf{Mitigation Strategies for Negative Impact on SE}
\end{itemize}
Studies have employed various strategies to mitigate the negative impact of human aspects on SE and these studies have proposed several strategies for mitigating these effects.
These strategies include managerial interventions focused on coordination and partnering to reduce perception gaps and conflicts, while improving project performance, interactions, and communication. Establishing coordination channels and building collaborative relationships can enhance stakeholder collaboration. Additionally, adopting organisational patterns facilitates improved communication and understanding between customers and the development team. Creating an infrastructure for managing the feedback in user reviews helps to streamline the developer-user communication via app reviews, resulting in an enhanced user experience. Further clear articulation of assumptions, provision of meeting notices, and adherence to communication principles are essential for effective communication. Maintaining proper communication with all stakeholders from the outset promotes team commitment to effective planning.

Formal charters can be formulated to establish shared objectives and responsibilities of users and developers. Implementing organisational coordination techniques and partnering procedures before project commencement is beneficial for effective project management. Also potential conflicts and problem areas should be identified before project initiation. Achieving mutual understanding in terms of success definitions and measures enhances the commonality among project stakeholders. 
Conducting dedicated meetings for reviews, retrospectives, and pre-grooming creates opportunities for team members to address areas where they may have insufficient knowledge or understanding. Studies have also suggested implementing agile approaches such as scrum to overcome interpersonal and intrapersonal challenges.
Providing coaching and guidance for leadership roles, fostering in-house leadership with senior guidance, having dedicated feedback sessions and retrospectives, employing software to measure office/team mood, and frequent supervision of these reports by the leader are key strategies for solving lack of leadership issue.

\begin{boxA}
    \footnotesize
    \textbf{Answer to RQ3:}
    We classified the impact of human aspects into two major categories as impact on developer-user interactions and impact on SE. We identified  positive, negative, or mixed effects in each category. The majority of the studies identified positive effects. Effects on developer-user interactions were categorised into four groups: effects on developers, users and intermediaries, the developer-user relationship, and the body of knowledge. Considering the effects on SE, we identified five groups, covering stakeholders and relationships, organisation and software projects, developed systems, SE processes, and contributions to existing knowledge. We didn't observe any negative effects on the body of knowledge in both categories. We further identified strategies employed by primary studies to mitigate the negative impact of human aspects on developer-user interactions and SE.
\end{boxA}


\subsection{Limitations and Future Work Recommendations (RQ4)}

\subsubsection{Study Limitations}
We identified various limitations in all the primary studies, which fell into four main categories: study methodology, study evaluation, research focus, and study results/outcomes (see Table \ref{TAB:Study Categorisation by Limitations}). Some studies acknowledged multiple limitations, while others mentioned only one.

\begin{itemize}
    \item \textbf{Limitations related to Methodology}
\end{itemize}
We identified three subcategories among the limitations related to methodology: study participants; tools, techniques, instruments; and study design. Most studies (35 studies) identified one of more issues under this category.

\begin{itemize}
    \item[--] \emph{Limitations related to Study Participants}
\end{itemize}
Studies in this category reported various limitations related to expertise/ experience of participants, demographics of participants, participant bias, number of participants/sample size, and representativeness of the sample. In total, we identified 17 studies reporting this limitation. For example, in IEEE06, the study was based on a customer-driven course at the Computer Science department of the Norwegian University of Science and Technology. The researchers selected a sample of SE students from this course, but the sample size was quite small, and the study was based on only one course. Therefore, the authors acknowledged that a larger sample size would be needed to generalise the results.

\begin{itemize}
    \item[--] \emph{Limitations related to used Tools/Techniques/Instruments}
\end{itemize}

We identified eight studies with limitations related to tools and techniques used, and their data collection instruments such as surveys and interviews. ACM01 discussed how developers and users express emotions and politeness by analysing a GitHub data set. The used emotion detection tool was trained with Jira comments, while the politeness tool was trained using Stack Overflow data. They have identified this variability as a potential threat to the reliability of the emotions and politeness tools applied in the research.

\begin{table} [htbp]
\centering
\scriptsize
\caption{Study Categorisation by Limitations}
\label{TAB:Study Categorisation by Limitations}
\begin{tabular}{P{0.2\linewidth} P{0.25\linewidth} P{0.5\linewidth}}
    \toprule
    \textbf{Main Category} & \textbf{Sub Category} & \textbf{Study IDs} \\
    \midrule
    
     \multirow{3}{\linewidth}{Limitations related to methodology} & Limitations in study design & CHASE01, SBB02, SBB03, SBB15, SBB14, SBF01, IEEE03, SBF02, SBF07, SBF06 \\
    
     & Limitations related to study participants & IEEE01, IEEE02, SPR01, SPR02, SBB12, IEEE06, CHASE02, SBB01, SBB02, SBB05, SBB06, SBF01, IEEE03, SBF02, SBF06, SBF08, SBF11\\
     
     & Limitations in used tools/ techniques/instruments & WILEY01, ACM01, SBB01, SBB04, SPR02, SBB08, SBF11, SBB12 \\

     & &\\
     

    Limitations in study evaluation &  & IEEE01, IEEE02, IEEE04, WILEY01, SD01, SBB12, IEEE05, IEEE06, CHASE02, SBB01, SBB07, SBB08, SBB11, SBB19, SBB15, SBB16, SBF03, SBB18, SBF02, SBF04, SBF07, SPR03, SBF05, SBF06, SBF10, SBF11, SBB06, IEEE03, SBB03, SPR04, SBF09, SBB17, SBF12 \\
    & &\\


    Limitations related to research focus & & SPR01, IEEE05, SBB02, SBB06, SBB15, SBB14 \\
    & &\\


    Limitations in study results/outcomes & & SPR01, SPR02, SBB01, SBB02, SBB09, SBB10, SBB13 \\
    \bottomrule

\end{tabular}
\end{table}

\begin{itemize}
    \item[--] \emph{Limitations related to Study Design}
\end{itemize}
A total of ten studies reported limitations related to study design, data analysis, and other methodology related issues. For example, SBB15 explored communication between developers and users via Apple App Store reviews. 
The researchers collected data based on a list of the most downloaded applications. This was identified as a limitation in study design, as it may have influenced the relationship between number of downloads and other variables such as ratings. This study made two assumptions during data analysis. Firstly, researchers have assumed missing publishing dates of reviews by using the dates of other reviews. Secondly, they adopted manual analysis to analyse feedback content. These assumptions were identified as data analysis limitations that may affect the internal validity of the results.

\begin{itemize}
    \item \textbf{Limitations in Study Evaluation}
\end{itemize}
Thirty three studies report various issues relating to the evaluation of their study. For example, IEEE01 is an study about enhancing developer-user communication. 
They emphasised the need for enhanced developer-user communication, especially in translating user needs into system requirements in large-scale IT projects using traditional development methods. Expert feedback was used to evaluate the study outcomes. However experts' experiences of communication setups in large-scale IT projects may vary based on factors such as application domain and nature of project. This variability could impact the generalisation of the results.

\begin{itemize}
    \item \textbf{Limitations related to Research Focus}
\end{itemize}
Six studies report limitations related to the focus and scope of the research, as well as those related to the human aspect(s) they studied.  For example, in SBB14, researchers studied the dynamic nature of mobile app reviews, specifically the dialogue that takes place between users and developers in the Google Play Store apps. However, they have only considered free apps, which is a limitation in the scope of the research. 

\begin{itemize}
    \item \textbf{Limitations in Study Results/Outcomes}
\end{itemize}
We identified seven studies with limitations related to study results, such as applicability, reliability, representativeness, bias and implications. SBB09 explored the impact of customer communication and feedback on defect rates. 
The researchers have indicated that the findings are more applicable to traditional software development as its iterations are relatively long compared to agile development. However, they have indicated that even agile development may benefit from the rich communication mechanisms proposed in this paper. This was identified as a limitation in the applicability of study results.

\subsubsection{Future Work Recommendations}
Most of the primary studies have suggested future work directions (36 studies). We identified four categories of future work: implement the findings in practice, extend research based on current findings, further explore a research area, and develop new model/platform/method or upgrading existing models (see Table \ref{TAB:Study Categorisation by Future Work}).

\begin{itemize}
    \item \textbf{Implementing Study Findings in Practice}
\end{itemize}
Four studies proposed implementing their preliminary findings in industrial settings. For example, SBB19 explored reasons for difficulties in developer-user communication. They have identified the cultural differences as the reason for this problem.
They suggested a list of recommendations to reduce the `people costs' of `insensitivity' and `obstructiveness'. These recommendations can be implemented in practice to mitigate the communication difficulties caused by cultural differences.

\begin{table} [htbp]
\centering
\scriptsize
\caption{Study Categorisation by Future Work}
\label{TAB:Study Categorisation by Future Work}
\begin{tabular}{P{0.36\linewidth} P{0.55\linewidth}}
    \toprule
    \textbf{Category} & \textbf{Study IDs} \\
    \midrule
    
    Implement findings in practice & SBB12, SBB19, SBB15, SBB16 \\
    
    Extending research based on current findings & WILEY01, SPR01, SPR02, IEEE06, CHASE01, CHASE02, SBB01, SBB05, SBB07, SBB08, SBB13, SBB14, SBF01, SBF02, SPR03, SBF05, SBF06, SBF10, SBF08, SBF11 \\
    
    Further exploring a research area & IEEE04, ACM01, WILEY01, SPR01, SBB12, IEEE06, SBB01, SBB06, SBB09, SBB11, SBB15, SBB14, SBB18, IEEE03, SBF07, SBF09, SBF12 \\
    
    Developing new model/platform/ method or upgrading existing models & IEEE01, WILEY01, SBB06, SBB15, SBB18, SBF04, SBF06 \\
    \bottomrule

\end{tabular}
\end{table}

\begin{itemize}
    \item \textbf{Extending the Research based on Current Findings}
\end{itemize}
Twenty primary studies proposed various generalisation related future work such as using a bigger sample, conducting more experiments/ evaluations, or improving the proposed solution. In SPR02, researchers explored UDC in large scale IT projects, gaps in communication and how to overcome them. They found that direct UDC is limited, and there is no commonly used method for this in design and implementation phases. The researchers stated their plans for future work, which involve proposing a method to support UDC in large-scale IT projects and evaluating the implementation feasibility through a case study to measure the benefits of this new method.

\begin{itemize}
    \item \textbf{Further Exploring a Research Area}
\end{itemize}
Seventeen studies proposed future work related to conducting more research on the same area, related area, or a new area. For example, SBB12 is focused on individual, group and organisation human aspects related to challenges in SE. It emphasised the importance of considering personality of: individuals when solving complex SE problems, recipients when communicating verbally, and employees when coping with organisational changes. It has pointed out that human behaviour is too complex to be understood just using one group of human aspects and suggested to explore practitioners' perspective to understand matters that pose more importance.


\begin{itemize}
    \item \textbf{Developing New Models or Upgrading Existing Models}
\end{itemize}
Seven studies proposed future work related to the development of new models, platforms, methods or upgrading existing models or platforms. SBF04 described a project named CloudTeams. CloudTeams aims to bridge the communication and collaboration gap between developers and end-users during software development. 
The future work for this study is centred on building the CloudTeams platform over the next two years.

\begin{boxA}
    \footnotesize
    \textbf{Answer to RQ4:}
    We found limitations in all primary studies included in this SLR, categorised into four main groups: study methodology, study evaluation, research focus, and study results. Among the limitations related to methodology, we identified subcategories involving study participants, techniques/instruments, and study design, with a majority of studies in this category. Regarding future work reported in the studies, we categorised them into implementing findings in practice, extending research based on current findings, further exploring a research area, and developing new models/platforms/methods or upgrading existing ones. Most studies mentioned future work related to further exploring a research area.
\end{boxA}

\section{Discussion} \label{SEC:Discussion}

\subsection{Implications for Research}
From our analysis of the limitations and future work proposed in the primary studies, we have identified some areas where more research could be undertaken. Based on this analysis, we offer some recommendations for future research to address these gaps.

\faIcon{graduation-cap} \textbf{Human aspects research related gaps:} Among the selected primary studies, there are only a few studies (1-2 studies) focusing on empathy, personality, skills, coordination, attitude, and competence related to developer-user interactions. Exploring the impact of these under-researched human aspects and determining the optimal way to integrate them into developer-user interactions in SE activities may enhance these interactions \cite{grundy2021impact, grundy2021addressing}. These human aspects can be studied individually as well as in combination with other human aspects. These future studies may beneficial in exploring positive and negative impacts of these under-researched human aspects on developer-user interactions. Further they may assist in understanding the ways to capture these human aspects in real-world, and determining the practical ways of improving SE by incorporating their impact.

\faIcon{graduation-cap} \textbf{Empathy in developer-user interactions:} Empathy was somewhat surprisingly found to be an under-researched area in terms of developer-user interactions in SE, as we found only two primary studies (SPR03, SBF08) regarding empathy. Both these studies were focused on improving developers’ understanding of usability. Empathy has been considered vital for fostering human connections in different disciplines \cite{hojat2016empathy}. However in SE, the role of empathy in developer-user interactions, effects of empathy in these interactions and the ways of improving SE outcomes and processes by leveraging empathy has not been studied so far. 

\faIcon{graduation-cap} \textbf{Research on developer-user personality:} There have been numerous studies investigating the impact of personality on SE tasks \cite{cruz2015forty}. Previous research has also identified SE activities that are influenced by the personality traits of SE professionals \cite{barroso2017personality}. However, our selected primary studies had only one study which emphasised the importance of considering personality in different SE challenges related to developer-user interactions (SBB12). Therefore, it seems useful to understand how developer and user personalities influence their interactions such as developer-user communication, decision making, and conflict resolution between developers and users. 

\faIcon{graduation-cap} \textbf{Research on attitude:} We found that the area of attitude is under-explored, with only one primary study (IEEE02) addressing this aspect among the papers included in this SLR. This study identified user-developer attitude and user attitude towards the system as critical factors for user engagement success in system development. However, the broader dynamics of how positive or negative attitudes impact the willingness of users and developers to engage with each other, and the evolution of attitudes throughout the course of a software project, remain unexplored in the existing literature. Investigating these aspects could provide valuable insights into the intricate interplay of attitudes in the developer-user relationship.

\faIcon{graduation-cap} \textbf{Coordination related research:} Coordination is another under- represented developer-user interaction area. One selected primary study (SBB05) found coordination has a positive impact on project performance during the system development process. Likewise coordination may also be helpful in resolving developer-user conflicts and improving other SE practises. Therefore future research may provide valuable insights into the role of coordination in developer-user interaction by understanding effective coordination practices in resolving developer-user conflicts, and how developer-user coordination needs change over the software development life cycle. 
 
\faIcon{graduation-cap} \textbf{Competence related research:} Two primary studies related to competence were found and these emphasised the importance of user competence in user engagement with developers (IEEE02, SBB01). Similarly competence may also influence other aspects of developer-user interaction such as communication, understanding. Hence further studies may helpful in understanding the role of competence in fostering effective developer-user communication, reducing misunderstandings, and aligning expectations.


\faIcon{graduation-cap} \textbf{Research on the relationships among different human aspects:} This was identified as an area with limited studies to date. \emph{Communication and collaboration} and \emph{culture and communication} human aspects combinations have been studied in more than one primary study. However,  the other human aspect combinations were studied only once which opens several possibilities of future research areas. By far the majority of the selected primary studies (86\%) have not investigated the relationships among different human aspects. Thus, more research studies are required to identify the relationships among different human aspects, and their impact on developer-user interactions.
By exploring these relationships, researchers can provide further insights into how software teams can effectively leverage various human aspects to improve overall project success and collaboration.

    

\faIcon{graduation-cap} \textbf{Lack of a proper human aspects taxonomy:} Similar to prior studies, we identified another limitation during our analysis, which is a lack of agreed definitions in studies of `human aspects' \cite{hidellaarachchi2021slr}. While categorising the identified human aspects, we realised that there is only one preliminary taxonomy for end-user human aspects in SE \cite{grundy2021addressing}. We didn't find any widely agreed taxonomy developed for the human aspects of the developers in the SE context. However, we used this taxonomy for our classification as the human aspects discussed in these studies are common to both developers and users.
There is thus a need for conducting evaluations using the current taxonomies and extending them to cater to human aspects of other SE stakeholders (e.g., developers, testers). There may be a need for developing new taxonomies to better capture diverse human aspects of these other stakeholders.

\faIcon{graduation-cap} \textbf{Developing a taxonomy to classify the effects of human aspects:} There is no known comprehensive taxonomy to classify the effects of human aspects. Developing a taxonomy to classify the effects of human aspects could significantly enhance our understanding of the intricate dynamics at play in developer-user interactions. During our SLR, we observed that several other SLRs have explored the effects of individual human aspects as well as human aspects in general \cite{barroso2017personality, beecham2008motivation, mohanani2020coginitvebias, hidellaarachchi2021slr, soomro2016personality}. However, none of these studies employed a taxonomy to systematically discuss the identified effects of human aspects. 
Introducing a new taxonomy in future research endeavours may provide a more structured approach to analyse the impact of human aspects and offer valuable insights into how the field has evolved over time. By reclassifying primary papers based on this new taxonomy, researchers can uncover trends, patterns, and changes in our understanding of human aspects in SE.

\faIcon{graduation-cap} \textbf{Addressing evaluation limitations:} Many primary studies had limitations in their evaluations, which were identified either as limitations within the study itself or during our analysis of the study. Evaluation of most of the proposed tools, theories or guidelines in the primary studies are planned to be completed as future works. Many studies admitted issues related to self-reported information, threats to reliability, and ability to influence the direction of the discussion in semi-structured interviews. Many studies had small sample size, specific geographic focus, variation in participants' expertise, sample size concerns, and context-specificity. Therefore proposed solutions still remain to be tested and evaluated with real-world scenarios and generalised contexts.

\textcolor{black}{\faIcon{graduation-cap} \textbf{Exploring the correlation between query specificity and snowballing efficacy:} Future SLRs could focus on studying the correlation between the specificity of search queries employed in SLRs and the efficacy of snowballing techniques. This investigation could involve examining how the level of specificity in search queries influences the need for and impact of snowballing in different domains. By gaining insights into this correlation, researchers can develop tailored search strategies that optimise the effectiveness of snowballing, particularly in domains with varying degrees of specificity in research queries.}

\subsection{Implications for Practice}

From our analysis of the findings and recommendations the primary studies, we have identified practical implications of our research for industrial practice. By integrating these insights into our discussion, we aim to offer a more comprehensive understanding of how our research findings can be translated into actionable strategies and practices for practitioners in the field.

\faIcon{laptop} \textbf{Addressing developer-user perception gaps:} Research has identified several factors contributing to developer-user perception gaps including uncertainty, instability, and diversity of requirements (SPR01, SBB05, SBB06). It has been shown that top management support, collaboration, and effective communication are instrumental in reducing these perception gaps. In addition, studies have highlighted the negative impact of perception gaps on user participation and project performance. These studies have proposed strategies for software practitioners and project managers to mitigate these challenges. These include fostering collaborative relationships among stakeholders, formalising shared objectives and responsibilities through a formal charter, proactively identifying potential conflicts and problem areas before project initiation, and facilitating both scheduled and unscheduled group meetings between users and practitioners. Additionally, project managers are encouraged to develop their contacts, contracts, and relationships with users early in the project life cycle to foster a deeper understanding and alignment of objectives.

\faIcon{laptop} \textbf{Expressing emotions and politeness in developer-user communication:} Studies have delved into the nuances of how various roles in open-source software development express emotions and politeness by analysing GitHub projects (ACM01). The findings revealed notable distinctions between developers and users in their expressions of emotions and levels of politeness. Developers' comments tend to exhibit lower levels of politeness compared to users' comments, with developers receiving a higher degree of politeness in response to their comments than users. Further, developers received a higher level of love and joy in the comments on their issues, while users received more expressions of anger and sadness in theirs compared to developers. These findings suggest that while users often strive to maintain politeness and convey positive emotions, their efforts may not always be reciprocated. As a result, researchers have recommended that developers exercise greater caution and mindfulness in their communications with other contributors. By recognising these disparities, software organisations can implement strategies to foster a more respectful and supportive atmosphere. Creating guidelines or best practices for communication within development teams can help set expectations and promote a culture of mutual respect and understanding. Additionally, providing training or workshops on effective communication techniques and conflict resolution can empower developers to navigate interpersonal challenges more effectively.

\faIcon{laptop} \textbf{Enhancing developer-user communication:} Many studies have offered insights into improving communication between developers and users by addressing various aspects of their interaction. Some of these recommendations include addressing cultural differences (SBB19, SBF10, SBF07), using intermediaries to facilitate communication (IEEE05, SBF07), suggestions on communication media selection (SBB09), and enhancing communication in RE process (IEEE01, IEEE04, SBB17, SBF02). 
These insights offer valuable strategies for industry practitioners to enhance communication between developers and users.
Especially researchers have emphasised that cultural differences between users and developers are at the foundation of the communication gaps (SBB19, SBF07, SBF10). 
Some of the recommendations to bridge these gaps include avoiding specialist terminology and explaining key jargon, considering the use of language from different perspectives to eliminate ambiguity, providing advance notice of meetings and agenda details for adequate preparation, recognising the distinct purposes of technology use for business users versus developers, 
respecting values associated with different roles, and treating business users as clients by investing time in understanding their needs (SBB19, SBF10).

\faIcon{laptop} \textbf{Improving developers' understanding of usability via empathy:} Research suggests a strong link between developers' understanding of usability and their empathy toward users' needs (SBF08, SPR03). They found that engaging in user-based thinking aloud protocols during usability evaluation sessions enhances developers' empathy by exposing them to users' emotions, fostering emotional contagion and heightening empathy toward user needs. This calls for alternative corrective measures over traditional training approaches. For the software industry, these findings underscore the importance of implementing targeted strategies to improve product development processes and user satisfaction. Incorporating usability evaluation methods that encourage direct interaction with users, such as user-based thinking aloud protocols can foster a culture of empathy among developers, leading to software products that better meet user needs. Recognising emotional contagion as a driver of developers' empathy emphasises the need for positive emotional experiences throughout the development life cycle, encouraging user-centred design practices and empathetic communication channels between developers and end-users. Ultimately, embracing empathy-driven approaches to usability can enhance product quality, user experience, and customer satisfaction.

\faIcon{laptop} \textbf{Addressing developer-user understanding gaps:} Studies highlighted understanding gaps between developers and users as critical in IT system failures (SBB04). They found that using set theory and Venn diagrams in RE process to illustrate relationships and gaps, aids to resolve understanding gaps. Another study identified key challenges in developer-user understanding 
and they introduced some organisational patterns to address them. 
\textit{Customer-driven design pattern} emphasises clear initial requirements and various approaches for customer involvement in design. \textit{Customer involvement with the development team pattern} advocates for engaging expert customers in the development process to uncover hidden needs. \textit{Illustrating business processes pattern} aids in understanding organisational flows, while business process re-engineering streamlines processes for approval. \textit{Satisfying customers by results pattern} involves designing prototypes for early feedback. \textit{Building master trainers pattern} enhances customer capacity for long-term impact. These strategies enhance understanding between developers and users, ultimately improving project outcomes and user satisfaction.

\faIcon{laptop} \textbf{Addressing developer-user gaps in general:} The researchers have delineated disparities between IT practitioners and users, as well as variances within each group by introducing the gap model (SBB08). 
Their findings revealed that IT professionals felt more in control of technology, whereas users often experienced helplessness when systems failed to meet their expectations. 
They also found that training and involvement in IS development significantly influenced user attitudes, with technically trained or involved users exhibiting IT professional-like attributes. This underscores the role of education and involvement in shaping user perceptions and attitudes toward technology.
Another study proposed enhancements to traditional IT curricula to bridge the IT-user gap (SBB16). It suggested two key approaches for hybrid IT major and hybrid IT minor curricula. 
For the hybrid major, students are encouraged to specialise in certain areas and collaborate on projects requiring diverse expertise. This approach fosters collaboration and reduces the perception of needing to know everything. 
The hybrid minor option aims to train non-IT majors as effective liaisons between their work group and the IT department. One approach is to create an IT liaison major, combining business core with technical background. 
Another approach is to offer an end-user support minor, providing technical training in end-user development, web development, and business-oriented software. 



\section{Threats to Validity} \label{SEC:Threats to Validity}
Our SLR is subject to certain threats even though it adheres to the well-accepted SLR process introduced by Kitchenham et al. \cite{Kitchenham2007Guidelines} \cite{Kitchenham2004Procedures}. We discuss the threats associated with each stage of our SLR and the mitigation strategies used during our study.

\textit{Data source and search strategy.} 
Our primary search is not time-bound, but it only includes papers published until April 2023. Therefore, we may have overlooked relevant studies published after that date. Constraints on the number of search terms in digital databases also posed a challenge. To address these threats, we selected the most commonly used digital databases, used consistent search terms across all databases, refined search strings for each database, executed multiple search queries in one database due to search term restrictions, and employed an exhaustive snowballing approach. Also when selecting the search terms, ``human aspects" term and its synonyms were used without adding any individual human aspects to ensure our search is not biased towards any human aspect. Both primary and secondary search phases covered more than one thousand publications and resulted in a broad set of primary studies.

\textit{Study filtration.} 
Study filtration was initially performed by a single researcher, which may have led to the possibility of missing relevant studies. However, we followed a comprehensive selection process as explained in Section \ref{SEC:Filtering of the Papers}, including three screenings, clear inclusion/exclusion criteria, and cross-validation by multiple authors. Regular discussions and quality assessments were also part of the filtration process as described in Section \ref{SEC:Quality Assessment}.

\textit{Data extraction and synthesis.} 
Data extraction and synthesis were conducted by one researcher, potentially introducing bias. Additionally, some primary studies lacked sufficient details for extraction. To mitigate these threats, cross-validation was performed for data extraction as described in Section \ref{SEC:Data Extraction Strategy}, and primary studies with less than three pages were excluded during filtration. Data synthesis was discussed with other authors during regular fortnightly meetings to ensure accuracy.

\section{Conclusion} \label{SEC:Conclusion}
We conducted this SLR with the aim of identifying the current body of work in human aspects related to the interactions between software developers and users. We performed a comprehensive search on six digital databases and adopted an exhaustive backward and forward snowballing process. We followed a thorough filtration process comprising three screening phases and 46 papers were selected for data extraction. We synthesised data to answer our four research questions. We found key objectives for conducting these primary studies, and details regarding methodology and participants of each study. We identified human aspects studied in each primary study and categorised them based on a taxonomy of human aspects. We also discovered the relationships among human aspects and the impact of human aspects explored in the primary studies. We analysed the outcomes, evaluation methods, limitations and future work of each primary study. Using these identified limitations and future work, we framed the key research gaps and proposed recommendations for future research and practice. The findings of our SLR will be beneficial in three ways. First, we now have an overview of the positive and negative impacts of human aspects on the interactions between developers and users as well as SE in general. Second, our findings will help to identify ways to leverage the positive effects in developer-user interactions as well as address the negative effects of these human aspects by following the mitigation strategies. These insights will be useful for software practitioners to bridge the gap between developers and users. These will also be helpful in having effective interactions with users. Finally the proposed recommendations will be helpful for the research community for conducting further research on human aspects or research involving human participants.    

\section*{Acknowledgements}
Gunatilake, Grundy and Mueller are supported by ARC Laureate Fellowship FL190100035.

\appendix
\section{List of Included Papers} \label{Appendix:List of included papers}

\begin{scriptsize}\setstretch{0.9}
\setlength{\tabcolsep}{2pt}
\renewcommand{\refname}{}

\end{scriptsize}
\renewcommand{\refname}{References}

\section{Positive Effects of Human Aspects} \label{Appendix:Positive Effects of Human Aspects}

\begin{tiny}
\begin{center}
\begin{ThreePartTable}
\begin{TableNotes}
    \item *P: Positive, M: Mixed, UPI: User Participation and Involvement, Challenges: Interpersonal and Intrapersonal Challenges
\end{TableNotes}

\begin{longtable}{P{0.05\linewidth} P{0.09\linewidth} P{0.13\linewidth} P{0.3\linewidth} P{0.35\linewidth}}
    \caption{Categorisation of Positive Effects by Human Aspects (summarised)}
    \label{TAB:Categorisation of Positive Effects by Human Aspects}\\
    


    \toprule
    \textbf{Category} & \textbf{Human Aspect} & \textbf{Nature of Impact \& Paper IDs} & \textbf{Impact on User-Developer Interactions} & \textbf{Impact on SE} \\
    \midrule
    \endfirsthead
    
   \multicolumn{3}{r@{}}{{\tablename\ \thetable{} -- \textit{Continued from previous page}}} \\
    \midrule
    \textbf{Category} & \textbf{Human Aspect} & \textbf{Nature of Impact \& Paper IDs} & \textbf{Impact on User-Developer Interactions} & \textbf{Impact on SE}\\
    \midrule
    \endhead
    
    \bottomrule
    \insertTableNotes
    \endfoot
    
    \bottomrule
    \insertTableNotes
    \endlastfoot
    

     \multirow{8}{*}{\rotatebox[origin=c]{90} {\parbox[c]{0.8cm}{Individual}}} & Empathy & *P:SPR03, SBF08  & Higher level of developer empathy towards users. & Increased system usability, Enhanced understanding of usability, Resource savings.\\
    
     & Motivation  & \makecell[tl]{P:IEEE02, \\SBB12, SBB01.\\ *M:CHASE02} & Developer empowerment, Improved developer motivation.  & Improved project success, Improved user commitment to the project. \\
     
     & Perception & \makecell[tl]{P:SBF06. \\M:SBB05} & Reduced perception gap, Improved client cooperation. & Successful client involvement in software projects. \\
     
     & Emotions & \makecell[tl]{P:ACM01. \\M:IEEE04}  & Quality developer-user relationship, Increased customer satisfaction. & Increasing productivity, Contributing to research of human factors, Improved requirement quality.  \\
     
     & Personality & P:SBB12 &  - & Resolving complex SE problems by acknowledging individual personality and organisational culture.  \\
     
     & Attitude & P:IEEE02 & - & Better understanding on the impact of attitude on engagement.  \\
     
     & Cognitive Style & \makecell[tl]{P:IEEE01, \\SBB04.\\ M:IEEE04, \\WILEY01.} & Reduced understanding gap between customers \& developers. & Increased UPI, More democratic organisational culture, Improved understanding of problem domain, Quality requirements.  \\
     
     & Competence & P:IEEE02, SBB01 & - & Better understanding of the impact of competence on engagement, Improved user commitment to project team.  \\
   
     \multirow{4}{*}{\rotatebox[origin=c]{90} {\parbox[c]{3cm}{Skill, Experiential or Environmental- influenced}}} & Human Values & \makecell[tl]{P: ACM01, \\SBB12. \\M: IEEE04} & Communication \& trust relationships between users \& developers. & Enhanced understanding on developer thought process, Contributing to research of human factors, Improved customer relations.  \\
    
     & Knowledge/ Education & \makecell[tl]{P: SBB04. \\M: IEEE04, \\WILEY01} & Increased knowledge sharing between employees, Effective project requirements definition. & Quality requirements, Increased customer satisfaction, Leading to better systems.  \\
     
     & Skills & M: WILEY01 & - & Supports successful implementation of the SPI efforts.  \\
     
     & Performance & \makecell[tl]{P: SBB01. \\M: SBB02, SBB05} & - & Increased project performance. \\
    
     \midrule
     
     \multirow{6}{*}{\rotatebox[origin=c]{90} {\parbox[c]{2cm}{Group Related}}} & Communication & \makecell[tl]{P: CHASE01, \\ACM01, SD01, \\SPR02, IEEE05, \\IEEE01, SBB09, \\SBB10, SBB13, \\SBB15, SBB14, \\SBB18, SBF01, \\SBF02 SBF04, \\SBF05, SBF10, \\IEEE02, SBB12\\ M: SBB05, \\SBF07, IEEE04} & Enable a richer communication, Improved developer understanding on user needs, Promotes users' positive attitude towards system, Enables effective use the systems, Facilitate better information flow for users, Increased efficiency of work, Improved opportunities for learning, Improved buy-in and ownership, Improved productivity. & 
     Improved SW \& data quality, Increased system success, Increased user satisfaction, Increased team productivity \& satisfaction, Better UX, Improved system usability, Increased project performance, Contributing to human factors research, Encouraging users to reflect on their technology usage, Reduced defect rate. \\
    
     & Collaboration & \makecell[tl]{P:SD01, IEEE06, \\SBB07, SBB10, \\SBF03, SBF04, \\SBF05, IEEE02, \\SBB12, SBB01\\  M:SBB02,SBB05, \\WILEY01} & Increased collaboration, Improved Buy-in \& ownership, Better developer-customer understanding, Reduced conflicts, Increased developer appreciation, Improved awareness on user participation \& collaboration. & Improved system usability, Assisting \& guiding decision making, Increased project performance, Increased user satisfaction, Reduced defect rate, Efficient use of resources, Improved partnership, Encouraging startup formation.  \\
     
     & Culture & \makecell[tl]{P:SBF10, IEEE02, \\SBB12, SBB04 \\M: SBF07} & Improved understanding of developer-user cultural differences, Improved communication \& interaction, Reduced difficulties, Quality requirements. & Improved system success, Better understanding on cultural impact, Collaborative customer relationship, Leading to better systems, Improved awareness of communications strategies, Resolution of complex SE problems.  \\
     
     & Coordination & M:SBB05 & Increased communication \& coordination due to collaboration.  & Increased project performance.  \\
     
     & *Challenges & \makecell[tl]{P:IEEE02 \\M:SBB02, \\SBB08} & Reduced developer-user conflicts, Increased awareness of developer-user differences, Improved communication \& understanding. & Reduced developer-user conflicts, Increased project success, Better understanding of the impact of challenges on engagement.\\
     
     & Engagement & \makecell[tl]{P:IEEE02,\\SBB01} & Increased UPI, Improved involvement of leadership \& employees, Improved developer-user collaboration. & Increased system success, Successful implementation of SPI efforts, Improved project performance. \\

\end{longtable}
\end{ThreePartTable}
\end{center}
\end{tiny}
\section{Negative Effects of Human Aspects} \label{Appendix:Negative Effects of Human Aspects}

\begin{tiny}
\begin{center}
\begin{ThreePartTable}
\begin{TableNotes}
    \item *N: Negative, M: Mixed, Challenges: Interpersonal and Intrapersonal Challenges
\end{TableNotes}

\begin{longtable}{P{0.05\linewidth} P{0.09\linewidth} P{0.13\linewidth} P{0.3\linewidth} P{0.35\linewidth}}
    \caption{Categorisation of Negative Effects by Human Aspects (summarised)}
    \label{TAB:Categorisation of Negative Effects by Human Aspects}\\
    


    \toprule
    \textbf{Category} & \textbf{Human Aspect} & \textbf{Nature of Impact \& Paper IDs} & \textbf{Impact on User-Developer Interactions} & \textbf{Impact on SE} \\
    \midrule
    \endfirsthead
    
   \multicolumn{3}{r@{}}{{\tablename\ \thetable{} -- \textit{Continued from previous page}}} \\
    \midrule
    \textbf{Category} & \textbf{Human Aspect} & \textbf{Nature of Impact \& Paper IDs} & \textbf{Impact on User-Developer Interactions} & \textbf{Impact on SE}\\
    \midrule
    \endhead
    
    \bottomrule
    \insertTableNotes
    \endfoot
    
    \bottomrule
    \insertTableNotes
    \endlastfoot
    

     \multirow{4}{*}{\rotatebox[origin=c]{90} {\parbox[c]{2cm}{Individual}}} & Motivation  & \makecell[tl]{*M:CHASE02} & Developer distrust in customer interactions, Customers feeling a lack of control, Weakened developer-customer interactions. & Customer frustration, Weakened developer - customer interactions.\\
     
     & Perception & \makecell[tl]{*N:SPR01, \\SBB06, SBB16. \\M:SBB05}  & Reduced user involvement, Reduced communication \& collaboration, Increased perception gaps. & Requirements uncertainty, Reduced project performance, Increased project failures. \\
     
     & Emotions & \makecell[tl]{N:SBB16. \\M:IEEE04}  & Limited interaction, Limited developer - user communication and trust. & Poor quality requirements, Insufficient domain knowledge, Low customer satisfaction, Failed software projects.  \\
     
     & Cognitive Style & \makecell[tl]{N:SBB16.\\ M:IEEE04, \\WILEY01.} & Lack of trust, Strained developer - user relationship, Failed communication. & Risks to project success, Failed software projects, Poorly coordinated development efforts.  \\
   
     \midrule
     \multirow{4}{*}{\rotatebox[origin=c]{90} {\parbox[c]{3cm}{Skill, Experiential or Environmental- influenced}}} & Human Values & \makecell[tl]{M: IEEE04} & Limited user interaction, Limited developer - user communication, Lack of trust. & Poor quality requirements, Insufficient domain knowledge, Low customer satisfaction.  \\
    
     & Knowledge/ Education & \makecell[tl]{M: IEEE04, \\WILEY01} & Limited user interaction, Limited developer-user communication, Lack of trust. & Poor quality requirements, Risks to project success, Failed SPI efforts.  \\
     
     & Skills & M: WILEY01 & - & Failed SPI efforts.  \\
     
     & Performance & \makecell[tl]{N: SBB06. \\M: SBB02, SBB05} & Developer - user conflicts, Reduced user - developer communication \& collaboration, Increased perception gaps, Lack of stakeholder understanding. & Increased project failures, Requirements uncertainty, Project estimation difficulties. \\
    
     \midrule
     
     \multirow{6}{*}{\rotatebox[origin=c]{90} {\parbox[c]{2cm}{Group Related}}} & Communication & \makecell[tl]{N: SPR04, \\SBB06, SBB11, \\SBB16, SBB19, \\SBB17, SBF09, \\SBF12\\ M: IEEE04, \\SBB05, SBF07} & Lack of understanding among stakeholders, Strained developer - user relationship, Failed communication, Cultural gap between stakeholders, Misalignment of stakeholders' objectives. & Inability of systems to meet business needs, Failed software projects, Poorly coordinated development efforts, Customer resistance toward software projects. \\
    
     & Collaboration & \makecell[tl]{N:SBB16\\  M:WILEY01, \\SBB02,SBB05} & Developer - user conflicts, Reduced user involvement, Reduced user - developer communication \& collaboration, Increased perception gaps, Strained developer - user relationship. & Failed SPI efforts, Increased project performance estimation difficulty, Reduced user - developer collaboration, Requirements uncertainty, Failed software projects.  \\
     
     & Culture & \makecell[tl]{N: SBB16, SBB19 \\M: SBF07} & Poor developer - user communication, Increased resistance, Strained developer - user relationship, Misalignment of stakeholders' objectives. & Inability of systems to meet business needs, Reduced usage of systems, Excessive maintenance, Failed software projects, Requirements uncertainty.  \\
     
     & Coordination & M:SBB05 & Reduced user involvement, Reduced user - developer communication \& collaboration, Increased developer - user perception gaps.  & Reduced project performance, Increased project failures, Requirements uncertainty.  \\
     
     & *Challenges & \makecell[tl]{N:IEEE03, SBB03, \\SBB16, SBF11 \\M:SBB02, SBB08} & Distant developer - user relationship, Developers' unawareness about customer needs, Negative user attitude towards technology, Developer demotivation, Lack of Empathy between users and developers. & Users feeling threatened by systems, Lack of user feedback, Obstacles in system development, Negating system success, Lack of confidence in system, Reduced team cohesion, Inadequate task analysis, Slow decision making, Reduced developer performance.\\
     
     & Engagement & \makecell[tl]{M:WILEY01} & - & Failed SPI efforts in SME software development companies.  \\

\end{longtable}
\end{ThreePartTable}
\end{center}
\end{tiny}

\bibliographystyle{elsarticle-num} 
\bibliography{References}





\end{document}